\documentclass[11pt,dvips]{article}
\usepackage{graphicx}
\usepackage{amsmath}
\usepackage{amssymb}
\usepackage{latexsym}
\usepackage{color}
\begin{document}
\thispagestyle{empty}
\begin{center}

\vspace{1.8cm}

 {\bf {\large Multipartite quantum correlations in even and odd  spin coherent states} }\\

\vspace{1.5cm}

{\bf M. Daoud}$^{a}${\footnote { email: {\sf
m$_{-}$daoud@hotmail.com}}}, {\bf R. Ahl Laamara}$^{b,c}$ {\footnote
{ email: {\sf ahllaamara@gmail.com}}} and {\bf W. Kaydi}$^{b}$
{\footnote { email: {\sf kaydi.smp@gmail.com}}}

\vspace{0.5cm}
$^{a}${\it Department of Physics , Faculty of Sciences, University Ibnou Zohr,\\
 Agadir ,
Morocco}\\[1em]

$^{b}${\it LPHE-Modeling and Simulation, Faculty  of Sciences,
University
Mohammed V,\\ Rabat, Morocco}\\[1em]

$^{c}${\it Centre of Physics and Mathematics,
CPM, CNESTEN,\\ Rabat, Morocco}\\[1em]

\vspace{3cm} {\bf Abstract}
\end{center}
\baselineskip=18pt
\medskip

\noindent The key ingredient of the approach, presented in this
paper, is the factorization property of $SU(2)$ coherent states upon
splitting or decay of a quantum spin system. In this picture, the
even and odd spin coherent states are viewed as comprising two,
three or more spin subsystems. From this perspective, we investigate
the multipartite quantum correlations defined as the sum of the
correlations of all possible bi-partitions. The pairwise quantum
correlations are quantified by entanglement of formation and quantum
discord.  A special attention is devoted to tripartite splitting
schemes. We explicitly derive the sum of entanglement of formation for
all possible bi-partitions.  It coincides with the sum of all possible
pairwise quantum discord. The conservation relation between the distribution
of entanglement of formation and quantum discord, in the tripartite
splitting scheme, is discussed. We  show that the
entanglement of formation and quantum discord possess the monogamy
property for even spin coherent states, contrarily  to odd ones
which violate the monogamy relation when the the overlap of the
coherent states approaches the unity.

\newpage
\section{Introduction}

The characterization of nonclassical correlations and nonlocal
correlations constitutes one of the main issues intensively
investigated in the field of quantum information science.
The primary goal  is to provide the best way to understand the
differences between quantum and classical physics.  Quantum correlations
constitute a relevant resource to manage information in several
ways \cite{Horodecki,Guhne,Modi}. Different forms
of measures to quantify the degree of quantumness in a multipartite
quantum system were introduced. In particular, entanglement of formation
has been successfully  employed in this sense. However, this measure does
not account for all nonclassical aspects of  correlations and
unentangled mixed states can possess quantum correlations. In this
respect,  other measures beyond entanglement were proposed in the
literature like for instance quantum discord introduced in
\cite{Ollivier-PRL88-2001,Vedral-et-al}. It is defined as the difference between
the total correlation and classical correlation present in a bipartite system. The quantum discord coincides with entanglement for pure
 states. For mixed states, the explicit evaluation of quantum discord involves an optimization
procedure which is in general a difficult task to achieve. There are
few two qubit systems
\cite{Luo,Ali,Shi1,Girolami,Shi2,Rachid1,Rachid2} for which
analytical results were obtained. To overcome the difficulty in
evaluating analytically quantum discord, a geometric method  was
introduced in \cite{Dakic2010}. Nowadays, entanglement of formation
\cite{Wootters98}, quantum discord
\cite{Ollivier-PRL88-2001,Vedral-et-al} and its geometric variant
\cite{Dakic2010} are typical examples of measures commonly used to
decide about the presence of quantum correlations between two
different parts composing a bipartite  quantum system.\\

In the recent years, the efforts in identifying and quantifying
quantum correlations were extended to correlated nonorthogonal
states as for example Glauber coherent states, $SU(2)$ and $SU(1,1)$
coherent states \cite{Sanders,Sanders2} (for a review see
\cite{Sanders3}). Subsequently,  many works have been devoted to
investigate their role    in quantum cryptography \cite{Crypto2},
quantum information processing \cite{Qip} and quantum computing
\cite{Bartlett,Jeong,Ralph}. This is mainly motivated by the
possibility to encode quantum information in continuous variables
\cite{Lloyd99}. For example,  the even and odd Glauber coherent
states, termed also Shr\"odinger cat states, can be considered as
basis states of a logical qubit \cite{Cochrane,Oliveira} and
provides a practical  way to implement experimentally optical
quantum systems useful
for quantum information.\\

In other hand, the structure of multipartite quantum systems is a
complicated and challenging subject that trigged off  a lot of
interest during the last decade (see \cite{Modi} and references
therein). In this paper, we shall strictly focus on the study of
quantum correlations present in odd and even $SU(2)$ coherent state.
In fact, by considering the property according to which a spin-$j$
coherent state $ |j, \eta \rangle$ can be factorized as a tensorial
product of two $SU(2)$ coherent states $ |j_1, \eta \rangle$ and $
|j_2, \eta \rangle$ with $(j = j_1 + j_2)$, it is possible to
construct a picture where even and odd spin coherent states might be
viewed as superpositions of two or more spin coherent  systems.
The idea of entanglement in a single particle, caused by quantum correlations
between  its
intrinsic degrees of freedom, was discussed in
\cite{Klyachko1,Klyachko2,Terra}. Consequently, it is seems natural
to assume that a odd or even spin-$j$ coherent state presents
quantum correlations between its intrinsic parts resulting from the
splitting of the spin$j$ into two or more subcomponents. In this
scheme, one can analyze the properties of multipartite quantum
correlations in many spin systems. The best way to approach this
question is the use of bipartite measures. This approach has the
advantage relying upon bipartite measures of entanglement of
formation  and quantum discord that are physically motivated and
analytically computable. Also, another important question emerging
in this context concerns the limitations of sharing quantum
correlations. Indeed, the distribution of quantum correlations among
the subsystems of a multipartite quantum system is constrained by
the so-called monogamy relation. It was firstly proposed by Coffman,
Kundo and Wootters in 2001 \cite{Coffman} in analyzing the
distribution of entanglement in a tripartite qubit system. Since
then, the monogamy relation  was extended to other measures of
quantum correlations. Unlike the squared concurrence \cite{Coffman},
the entanglement of formation does not satisfy the monogamy relation
\cite{Coffman} in a pure tripartite qubit system but it was reported
in \cite{Adesso2,Adesso3} that it can be satisfied in multimode
Gaussian states. Furthermore, quantum correlations, measured by
quantum discord, were shown to violate monogamy for some specific
quantum states \cite{Giorgi,Prabhu,Sudha, Allegra,Ren}. Now there
are many attempts to establish the conditions under which a given
quantum correlation measure is
monogamous or not. One may quote for instance the results obtained in \cite{Bruss} .\\

This paper is organized as follows. In Section 2 we give the
definitions of the bipartite measures: concurrence, entanglement of
formation and quantum discord. We also introduce the measure of
multipartite correlations in a given system as the sum of all
possible bipartite correlations. Section 3 concerns even and odd
spin coherent states. We especially discuss the decomposition
property of spin coherent states according to which they split in
multipartite spin or qubit systems. In Section 4, we derive the
explicit expressions of pairwise quantum correlations present in
even and odd spin coherent states decomposed in a pure bipartite
system. An appropriate qubit mapping is introduced. The results of
section 4 are extended in section 5 to the situation where the spin
coherent state splits in three spin sub-systems. A qubit mapping is
realized for all possible bi-partitions of the system. The total
amount of entanglement of formation is derived in Section 6.
Similarly, in Section 7, we explicitly evaluate the total amount of
quantum discord present in even and odd spin coherent states viewed
as a tripartite system.
The sum of pairwise quantum discord is evaluated. It  coincides with the
 total amount of bipartite entanglement of formation
in agreement with the result obtained in \cite{Z-H Ma}.
This result originates from the conservation relation between the distribution
of entanglement of formation and quantum discord proved in \cite{Fanchini1}.
Limitations to sharing entanglement of
formation as well as quantum discord are discussed. Some special
cases to corroborate our analysis  are numerically examined.
Concluding remarks close this paper.

\section{Quantum correlations}

The theoretical investigation of quantum correlations in a
multipartite quantum system is motivated by the recent experimental
progress in creating and manipulating  highly correlated spin
ensemble which provide experimentally accessible systems for quantum
information processing. In general the analysis of the properties of
quantum correlations in many spin systems is difficult. The simply
way to approach this problem is  the use of bipartite measures that
are explicitly computable such as entanglement of formation and
usual quantum discord. The definitions of each of these two measures
is presented here after. For an arbitrary tripartite state, the
quantum correlations present in the system can be computed by
considering all possible bipartite splits. The whole system can be
partitioned in two different ways. In  the first bi-partition
scheme, the system splits into two subsystems, one containing one
particle and the second comprising the two remaining particles. The
second bipartition is obtained by tracing out the degrees of freedom
of the third subsystem. In this picture, the total amount of quantum
correlations is given by the  sum of all possible bipartite quantum
correlations.

\subsection{Bipartite measures of entanglement of formation and quantum discord}
We shall first review briefly the concept of quantum discord
\cite{Ollivier-PRL88-2001,Vedral-et-al}. The total correlation is
usually quantified by the mutual information, usually expressed in
term of von Neumann entropy, as
\begin{equation}\label{def: mutual information}
    I(\rho_{AB})=S(\rho_A)+S(\rho_B)-S(\rho_{AB}),
\end{equation}
where $\rho_{AB}$ is the state of a bipartite quantum system
composed of the subsystems $A$ and $B$, the operator
$\rho_{A(B)}={\rm Tr}_{B(A)}(\rho_{AB})$ is the reduced state of
$A$($B$) and $S(\rho)$ is the von Neumann entropy of a quantum state
$\rho$. The mutual information $I(\rho_{AB})$ contains both quantum
and classical correlations. It decomposes  as
$$ I(\rho_{AB}) =  D(\rho_{AB}) +  C(\rho_{AB}),$$
and the quantum discord $D(\rho_{AB})$ is defined as the difference between the
total correlation $I(\rho_{AB})$  and the classical correlation
$C(\rho_{AB})$ present in the bipartite system $AB$. The classical
part $C(\rho_{AB})$ can be determined by optimizing local
measurement procedure as follows. Let us consider a von Neumann type
 measurement, on the subsystem $A$, belonging to the set one-dimensional projectors
  $\mathcal{M}=\{M_k\}$ with
 $\sum_k M_k= \mathbb{I} $. The von Neumann
measurement yields the statistical ensemble
$\{ p_{B,k} , \rho_{B,k}\}$ such that
$$\rho_{AB} \longrightarrow \frac{(M_k \otimes \mathbb{I})\rho_{AB}(M_k \otimes \mathbb{I})}{p_{B,k}}$$
where the measurement operation is written as \cite{Luo08}
\begin{eqnarray}
M_k = U \, \Pi_k \, U^\dagger \label{Eq:VNmsur}
\end{eqnarray}
with $\Pi_k = |k\rangle\langle k| ~ (k = 0,1)$ is the one
dimensional projector for subsystem $A$  along the computational
base $|k\rangle$,  $U \in SU(2)$ is a unitary operator and
$$ p_{B,k} = {\rm tr}  \bigg[ (M_k \otimes \mathbb{I})\rho_{AB}(M_k \otimes \mathbb{I}) \bigg]. $$
The amount of information acquired about particle $B$ is then given
by
$$S(\rho_B)-\sum_k ~p_{B,k} ~S(\rho_{B,k}),$$
which depends on measurement $\mathcal{M}$. To remove the
measurement dependence, a maximization over all possible
measurements is performed and the classical correlation writes
\begin{eqnarray}
    C(\rho_{AB})& =\max_{\mathcal{M}}
    \Big[S(\rho_B)-\sum_k ~p_{B,k} ~S(\rho_{B,k})\Big] \nonumber \\
    & =S(\rho_B) - \widetilde{S}_{\rm min}
      \label{def: classical correlation}
\end{eqnarray}
where $\widetilde{S}_{\rm min}$  denotes the minimal value of the
conditional  entropy
\begin{equation}\label{condit-entropy}
\widetilde{S} =  \sum_k ~p_{B,k} ~S(\rho_{B,k}).
\end{equation}
When optimization is taken over all perfect measurement, the quantum
discord is
\begin{equation} \label{def: discord}
    D(\rho_{AB})= I(\rho_{AB}) - C(\rho_{AB})
    =S(\rho_A)+\widetilde{S}_{\rm min}-S(\rho_{AB}).
\end{equation}
The explicit evaluation of quantum discord (\ref{def: discord})
requires the analytical computation of $\widetilde{S}_{\rm min}$.
This quantity was explicitly  derived only for few exceptional
two-qubit quantum states. One may quote for instance the results
obtained in \cite{Ali,Adesso1} (see also
\cite{Rachid1,Rachid2,Rachid3}). In this paper, we shall mainly
concerned with two-rank quantum states for which the minimization of
the conditional entropy (\ref{condit-entropy}) can be exactly
performed by purifying the density matrix $\rho_{AB}$ and making use
of Koashi-Winter relation \cite{Koachi-Winter} (see also
\cite{Shi}). This relation establishes the connection between the
classical correlation of a bipartite state $\rho_{AB} $ and the
entanglement of formation of its complement $\rho_{BC}$. Hereafter,
we discuss shortly this method. We assume that the density matrix
$\rho_{AB}$ has two non vanishing eigenvalues (two-rank matrix). It
decomposes as
\begin{eqnarray}
\rho_{AB} = \lambda_+ \vert \phi_+ \rangle_{AB}\langle \phi_+ \vert
+ \lambda_- \vert \phi_- \rangle_{AB} \langle \phi_- \vert
\end{eqnarray}
where  $\lambda_+$ and $\lambda_-$ are the eignevalues of $\rho_{AB}
$ and  the corresponding eigenstates are denoted by $\vert \phi_+
\rangle_{AB}$ and $\vert \phi_- \rangle_{AB}$ respectively. The
purification of the mixed state $\rho_{AB}$ is realized by attaching
a qubit $C$ to the two-qubit system $A$ and $B$. This yields
\begin{eqnarray}
\vert \phi \rangle_{ABC} = \sqrt{\lambda_+} \vert \phi_+
\rangle_{AB} \otimes \vert {\bf 0}  \rangle_{C} +  \sqrt{\lambda_-}
\vert \phi_- \rangle_{AB} \otimes \vert {\bf 1} \rangle_{C}
\end{eqnarray}
such that the whole system $ABC$ is described by the pure density
matrix $\rho_{ABC} = \vert \phi \rangle_{ABC} \langle \phi \vert $
from which one has  the bipartite densities $\rho_{AB} = {\rm Tr}_C
\rho_{ABC}$ and $\rho_{BC} = {\rm Tr}_A \rho_{ABC}$. According to
Koachi-Winter relation \cite{Koachi-Winter}, the minimal value of
the conditional entropy coincides with the entanglement of formation
of $\rho_{BC}$:
\begin{equation}
\widetilde{S}_{\rm min} = E(\rho_{BC})
\end{equation}
which is given by
\begin{equation}\label{stild-min}
 \widetilde{S}_{\rm min} = E(\rho_{BC}) = H(\frac{1}{2} + \frac{1}{2} \sqrt{1 - \vert {\cal C}(\rho_{BC})\vert^2})
\end{equation}
where  $H(x) = -x\log_{2} x  -(1-x)\log_{2} (1-x)$ is the binary
entropy function and ${\cal C}(\rho_{BC})$ is the concurrence of the
density matrix $\rho_{BC}$. We recall that for $\rho_{12}$ the density
matrix for a pair of qubits~$1$ and~$2$, which may be pure or mixed,
the concurrence is~\cite{Hil97}
\begin{equation}
{\cal C}_{12}=\max \left\{ \lambda _1-\lambda _2-\lambda _3-\lambda
_4,0\right\} \label{eq:c1}
\end{equation}
for~$\lambda_1\ge\lambda_2\ge\lambda_3\ge\lambda_4$ the square roots
of the eigenvalues of the "spin-flipped" density matrix
\begin{equation}
\varrho_{12}\equiv\rho_{12}(\sigma_y\otimes\sigma
_y)\rho_{12}^{\star}(\sigma_y\otimes \sigma_y), \label{eq:c2}
\end{equation}
where the star stands for complex conjugation in the basis $\{ \vert
00 \rangle, \vert 01 \rangle, \vert 10 \rangle, \vert 11 \rangle \}$
with the Pauli matrix is $\sigma_y = i \vert 1 \rangle \langle 0
\vert - i \vert 0 \rangle \langle 1 \vert$.
Nonzero concurrence
traduces the entanglement between the qubits
 1 and 2, otherwise they  are separable. Using the equations (\ref{def: discord}) and (\ref{stild-min}),  the quantum discord writes
as
\begin{equation}
D_{AB} \equiv D^{\rightarrow}_{AB} =  S_A - S_{AB} + E_{BC}.
\end{equation}
 In the same
manner, when the measurement is performed on the subsystem $B$, it
is simply verified that the quantum discord takes the form
\begin{equation}
D_{BA} \equiv D^{\leftarrow}_{AB} = S_B - S_{AB} + E_{AC}
\end{equation}
Notice that for a pure density matrix $\rho_{AB}$, the quantum
discord reduces to entanglement of formation $E(\rho_{AB})$.

\subsection{Multipartite quantum correlations}

The measure of multipartite quantum correlations constitutes an
important issue in the context of quantum information. Some attempts
to provide a precise way to quantify and characterize the genuine
multipartite correlations were discussed in the literature yielding
different approaches \cite{Z-H Ma,Okrasa,Chakrabarty,Rulli}. In particular,
Rulli and Sarandy \cite{Rulli} defined the multipartite measure of
quantum correlation as the maximum of the quantum correlations
existing between all possible bipartition of the multipartite
quantum system. In a similar way, Z-H Ma and coworkers \cite{Z-H Ma}
suggested a slightly different definition to quantify the global
multipartite quantum correlation. It is defined as the sum of the
correlations  in all possible bi-partitions. In this paper,
paralleling the treatment discussed in \cite{Z-H Ma},  we shall
quantify the global quantum correlations present in even and odd
spin coherent states as follows. For a tripartite spin coherent
states system $(j_1j_2j_3)$ arising from the decomposition  of a
spin-$j$ coherent state with $j = j_1+j_2+j_3$, the total amount of
quantum correlation is defined by
\begin{eqnarray}
Q(j_1,j_2,j_3) &=& \frac{1}{12} ( Q_{j_1j_2} +  Q_{j_2j_1} +
Q_{j_1j_3} + Q_{j_3j_1} + Q_{j_2j_3} + Q_{j_3j_2}  \nonumber \\ &+&
Q_{j_1( j_2j_3)} + Q_{(j_2j_3)j_1} + Q_{j_2( j_1j_3)} +
Q_{(j_1j_3)j_2} + Q_{j_3( j_1j_2)} + Q_{(j_1j_2)j_3}) \label{Qtotal}
\end{eqnarray}
where the bipartite measure $Q$ stands for entanglement of formation
or quantum discord. More details concerning the remarkable splitting
property of spin coherent states will be presented  in the next
section. In other hand, as we shall deal with tripartite quantum
states, it is natural to investigate the intriguing monogamy
relation of quantum correlation present in spin coherent coherent
states. The concept of monogamy can be introduced as follows. Let
$Q_{A\vert B}$ denotes the shared correlation $Q$ between $A$ and
$B$. Similarly, let us denote by $Q_{A\vert C}$ the measure of the
correlation between $A$ and $C$ and $Q_{A\vert BC}$ the correlation
shared between $A$ and the composite subsystem $BC$ comprising $B$
and $C$. The bipartite measure of correlations $Q$ is monogamous if
$Q_{A\vert BC}$ is greater  that the sum of $Q_{A\vert B}$ and
$Q_{A\vert C}$:
\begin{eqnarray}
Q_{A\vert BC} \geq Q_{A\vert B} + Q_{A\vert C}.
\end{eqnarray}
This inequality imposes severe limitations  to sharing quantum
correlations. The monogamy of entanglement of formation and quantum
discord in tripartite spin coherent states are examined in the
sections 6 and 7. It must be emphasized that the conditions under
which any measure of  quantum correlations that comprise and go beyond entanglement of formation  was
discussed by Fanchini et al in \cite{Fanchini2} for an arbitrary pure tripartite state. In particular, the authors developed
an elegant operational approach based on the  discrepancy  between classical and quantum correlations to set
up the constraints that any pure tripartite state must satisfy such that the entanglement of formation follow
the monogamy property.  This approach allows also to understand
the result obtained by Giorgi \cite{Giorgi} according to which the entanglement of formation and quantum discord obey the same monogamous relation.

\section{Spin coherent states as multi-qubit systems}

\subsection{Multi-qubit structure of Bloch coherent spin states}
An arbitrary spin system is described by the $su(2)$ algebra
generated by the operators $J_+, J_-$ and $J_3$ satisfying the
following structure relations
\begin{equation}
[J_{3} , J_{\pm}] = \pm J_{\pm}, \mbox{\hspace{1.0cm}} [J_{-} ,
J_{+}] = -2 J_{3} \ .   \label{4.2}
\end{equation}
The different irreducible representations classes of the group $SU
(2)$ are completely determined by the quantum angular momentum $j$
which may take integer or half integer values ( $j = {1\over 2}, 1,
\frac{3}{2}, \ldots$). The $(2j+1)$-dimensional Hilbert space is
spanned by the irreducible tensorial set $\{ \vert j , m \rangle, m
= -j , -j+1, \cdots, j-1, j\}$ characterizing the spin-$j$
representations of the group $SU(2)$. The standard $ SU(2)$ coherent
states are obtained by the action of an element of the coset space
$SU(2)/ U(1)$
\begin{equation}
 D_j(\xi) = \exp (\xi J_{+} - \xi^{\ast} J_{-}) \ ,
\end{equation}
on the extremal state $|j,-j\rangle$. This action  gives  the states
\begin{equation}
|j,\eta\rangle = D_j(\xi) |j,-j\rangle = \exp(\xi J_{+} - \xi^{\ast}
J_{-}) |j,-j\rangle = (1+|\eta|^{2})^{-j} \exp(\eta J_{+})
|j,-j\rangle \ ,   \label{4.7}
\end{equation}
where $\eta = (\xi/|\xi|)\tan |\xi|$. In the basis $\{ \vert j , m
\rangle\}$, they write
\begin{equation}\label{cs-spinj}
|j,\eta\rangle = (1+|\eta|^{2})^{-j} \sum_{m=-j}^{j} \left[
\frac{(2j)!}{(j+m)!(j-m)!} \right]^{1/2} \eta^{j+m} |j,m\rangle \ .
\end{equation}
They satisfy the resolution to identity property
\begin{equation}
\int d\mu(j,\eta) |j,\eta\rangle \langle j,\eta| = I \ ,
\mbox{\hspace{1.0cm}} d\mu(j,\eta) = \frac{2j+1}{\pi}
\frac{d^{2}\!\eta}{(1+|\eta|^{2})^{2}} \ .  \label{4.10}
\end{equation}
The spin coherent states are not orthogonal to each other:
\begin{equation}
\langle j,\eta_{1}|j,\eta_{2}\rangle = (1+|\eta_{1}|^{2})^{-j}
(1+|\eta_{2}|^{2})^{-j} (1 + \eta_{1}^{\ast} \eta_{2})^{2j} \ .
\label{4.9}
\end{equation}
The resolution to identity makes possible to expand an arbitrary
state in terms of the coherent states $|j,\eta\rangle$. In the
special case $j = \frac{1}{2}$, the spin coherent states
(\ref{cs-spinj}) reduce to
\begin{equation}
|  \eta \rangle = \frac{1}{\sqrt{1 + \bar\eta \eta}}
|\downarrow\rangle
           + \frac{\eta}{\sqrt{1 + \bar\eta \eta}} |\uparrow\rangle . \label{coh}
\end{equation}
Here and in the following $| \eta \rangle$ is short for the
spin-$\frac{1}{2}$ coherent state $|{\frac{1}{2}}, \eta\rangle$ with
$|\uparrow\rangle \equiv |\frac{1}{2},\frac{1}{2}\rangle$ and
$|\downarrow \rangle \equiv |\frac{1}{2} ,- \frac{1}{2}\rangle$). It
is important to notice that the  tensorial product of two  $SU(2)$
coherent states $ |j_1, \eta \rangle$ and $ |j_2, \eta \rangle$
produces a spin-$(j_1+j_2)$ coherent state labeled by the same
variable:
\begin{equation}\label{split}
 |j_1, \eta \rangle \otimes | j_2 , \eta \rangle
  =  (D_{j_1} \otimes D_{j_2})\, (|j_1,j_1\rangle \otimes |j_2,j_2\rangle)
  =  D_{j_1+j_2}\, |j_1+j_2,j_1+j_2\rangle
    \; = \; |j_1 + j_2, \eta \rangle .
\end{equation}
Only coherent states possess this remarkable property. It allows  to
write any spin-$j$ coherent states as a $2j$ tensorial product of
spin-$\frac{1}{2}$
 coherent states:
\begin{equation}
|j, \eta \rangle =  \left(|\eta\rangle \right)^{\otimes 2j}
            =   \left( \frac{1}{\sqrt{1 + \bar\eta \eta}}
|\downarrow\rangle
           + \frac{\eta}{\sqrt{1 + \bar\eta \eta}} |\uparrow\rangle \right)^{\otimes 2j} \nonumber
          =  (1 + \bar\eta \eta)^{-j}\sum_{m=-j}^{+j} {2 j \choose j + m}^{\frac{1}{2}}
         \eta^{j+m}
          |j,m\rangle,  \label{Coh}
\end{equation}
reflecting that a spin-$j$ coherent state may be viewed as a
multipartite state containing  $2j$ qubits.
\subsection{Even and odd coherent states}
The even and odd spin coherent states are defined by
\begin{equation}\label{ncs}
 \vert j, \eta , m \rangle  =  {\cal N}_m ( \vert j, \eta \rangle + e^{im\pi} \vert j, - \eta
 \rangle)
\end{equation}
where the integer $m \in \mathbb{Z}$ takes the values $m = 0 ~({\rm
mod}~2)$ and $m = 1~ ({\rm mod}~2)$. The normalization factor ${\cal
N}_m$ is
$$ {\cal N}_m = \big[ 2 + 2 p^{2j} \cos m \pi\big]^{-1/2}$$
where  $p$ denotes the overlap between the states $\vert \eta
\rangle$ and $\vert  -\eta \rangle$. It is given
\begin{equation}\label{overlap}
 p = \langle \eta \vert  - \eta \rangle = \frac{1 - \bar\eta \eta}{1 + \bar\eta \eta}.
\end{equation}
For  $ j = \frac{1}{2}$, the even and odd coherent states coincide
with $\vert \uparrow \rangle $ and $ \vert \downarrow \rangle$. They
can be identified with basis states for a logical qubit as $ \vert
0\rangle \rightarrow \vert \uparrow \rangle$ and $ \vert 1 \rangle
\rightarrow \vert \downarrow \rangle$. This line of reasoning can be
extended to higher spin values and provides scheme to encode
information in superpositions of arbitrary spin coherent states,
especially even and odd ones. Indeed, the states $ \vert j, \eta , 0
\rangle$ and $ \vert j, \eta , 1 \rangle$ define a two-dimensional
orthogonal basis and give a first possible encoding scheme. Thus,
one can identify the even state $ \vert j, \eta , 0 \rangle$ and the
odd state $ \vert j, \eta , 1 \rangle$ as basis of a logical qubit
as
$$ \vert
j, \eta , 0 \rangle \longrightarrow \vert  0 \rangle_j \qquad \vert
j, \eta , 1\rangle \longrightarrow \vert  1 \rangle_j. $$ Others
encoding schemes involving more qubits are also possible. They can
be realized using the factorization or the splitting property of
spin coherent states (\ref{split}). In fact, the states (\ref{ncs})
can be also expressed as
\begin{equation}\label{cs-2q}
 \vert j, \eta , m \rangle  =  {\cal N}_m ( \vert j_1, \eta \rangle\otimes\vert j_2, \eta \rangle +
 e^{im\pi}  \vert j_1, -\eta \rangle\otimes\vert j_2, -\eta \rangle)
\end{equation}
with $j = j_1+j_2$. They can rewritten as a two qubit states in the
basis
$$ \vert
j_i, \eta , 0 \rangle \longrightarrow \vert  0 \rangle_{j_i} \qquad
\vert j_i, \eta , 1\rangle \longrightarrow \vert  1 \rangle_{j_i},
\quad i=1,2.
$$
defined by means of  odd and even spin coherent associated with the
angular momenta $j_1$ and $j_2$. This construction is easily
generalizable to three and more qubits. In this manner,  the states
$ \vert j, \eta , m \rangle$ can be viewed as multipartite fermionic
coherent states:
\begin{equation}\label{cs-2j}
 \vert j, \eta , m \rangle  =  {\cal N}_m ( \left(|\eta \rangle \right)^{\otimes 2j} + e^{im\pi} \left(|-\eta \rangle \right)^{\otimes
 2j}).
\end{equation}
Furthermore, the logical qubits $ \vert j, \eta , 0 \rangle$ (even)
and $ \vert j, \eta , 1 \rangle$ (odd) spin coherent states behave
like a multipartite state of Greenberger-Horne-Zeilinger (${\rm
GHZ}$) type \cite{GHZ} in the asymptotic limit $p \rightarrow 0 $.
In this special  limiting case, the  states $|\eta \rangle $ and $|
-\eta \rangle $ approach orthogonality and an orthogonal basis can
be defined such that $\vert {\bf 0}\rangle\equiv \vert \eta \rangle$
and $\vert{\bf 1}\rangle \equiv \vert  -\eta \rangle$. Thus, the
state $\vert j , \eta, m \rangle$ becomes of ${\rm GHZ}$-type
\begin{equation}
\vert j , \eta, m \rangle \sim \vert {\rm GHZ}\rangle_{2j} = \frac
1{\sqrt{2}}(\vert {\bf 0}\rangle \otimes |{\bf 0}\rangle \otimes
        \cdots \otimes\vert {\bf 0}\rangle
    +e^{i m \pi}\vert {\bf 1}\rangle \otimes
    \vert {\bf 1}\rangle \otimes \cdots \otimes
\vert {\bf 1}\rangle).\label{GHZ}
\end{equation}
The second limiting case corresponds to the situation when $p
\rightarrow 1$ (or $ \eta \rightarrow 0$ ). In this case it is
simple to check that the state $\vert j , \eta, m = 0 ~({\rm mod}~
2) \rangle$ (\ref{cs-2j}) reduces to ground state of a collection of
$2j$ fermions
\begin{equation}
\vert j,  0 , 0 ~({\rm mod}~ 2) \rangle \sim  \vert \downarrow
\rangle \otimes\vert \downarrow \rangle \otimes \cdots \otimes \vert
\downarrow \rangle,
\end{equation}
and  the state $\vert j , \eta , 1 ~({\rm mod}~ 2) \rangle$  becomes
a multipartite state of $W$ type~\cite{Dur00}
\begin{equation}
\vert j,  0 ,  1 ~({\rm mod}~ 2) \rangle \sim \vert\text{\rm
W}\rangle_{2j}
    = \frac{1}{\sqrt{2j}}(\vert \uparrow \rangle \otimes\vert \downarrow \rangle \otimes \cdots\otimes
       \vert \downarrow \rangle  +\vert \downarrow \rangle \otimes\vert \uparrow\rangle \otimes\ldots\otimes \vert \downarrow\rangle
      +\cdots
   + \vert  \downarrow \rangle \otimes\vert \downarrow \rangle  \otimes \cdots\otimes \vert \uparrow\rangle)~.
\label{Wstate}
\end{equation}
The even spin coherent states  $\vert j, \eta , m = 0 ~({\rm mod}~2)
\rangle$ interpolate continuously between ${\rm GHZ}_{2j}$ states
$(p \rightarrow 0)$ and the completely separable state $\vert
\downarrow \rangle \otimes\vert \downarrow \rangle \otimes \cdots
\otimes \vert \downarrow \rangle$ $(p \rightarrow 1)$. In the odd
case, corresponding  to
 $\vert j, \eta , m = 1 ~({\rm mod}~2) \rangle$, we obtain states
interpolating between states of ${\rm GHZ}_{2j}$ type $(p
\rightarrow 0)$ and states of ${\rm W}_{2j}$ type $(p
\rightarrow 1)$.

\noindent The decomposition property (\ref{split}) provides us with
a picture where even and odd spin coherent states can be considered
as comprising multipartite spin subsystems. This is our main
motivation to investigate the quantum correlations present in a
single spin coherent state. This issue is discussed in what follows.

\section{Bipartite splitting and bipartite correlations}

In this section, we first discuss the bipartite splitting described
by  the equation (\ref{cs-2q}). In this scheme, the entire system
contains  two subsystems characterized by the angular momenta $j_1$
and $j_2$ such that $j = j_1 + j_2$. Accordingly, $(2j-1)$ possible
bipartite splitting are possible:
$$ j_1 = j - \frac{s}{2} \quad j_2 = \frac{s}{2} \quad s = 1, 2, \cdots, 2j-2, 2j-1,$$
and subsequently it is interesting to  compare the pairwise quantum
correlations  in each possible bipartite splitting.

\subsection{Bipartite entanglement of formation}

As discussed in the previous section, for each bipartition $s$ $( s
= 1, 2, \cdots, 2j-1)$, the coherent state $ \vert j, \eta , m
\rangle$ can be expressed as a state of two logical qubits. In this
sense, for each subsystem, an orthogonal basis $\{ \vert 0 \rangle_l
, \vert 1 \rangle_l\}$, with $ l = j_1$ or $j_2$, can be defined as
\begin{equation}\label{base0}
\vert 0 \rangle_l = \frac{ \vert l , \eta  \rangle +  \vert l ,
-\eta \rangle}{\sqrt{2(1 + p^{2l})}}
   \qquad \vert 1 \rangle_l = \frac{\vert l , \eta \rangle -  \vert l , -\eta
\rangle}{{\sqrt{2(1- p^{2l})}}}.
\end{equation}
The bipartite density matrix $\rho = \vert j, \eta , m\rangle \langle
j, \eta , m \vert$ is pure. In this situation, the quantum discord
for the pure state $\rho_{AB}\equiv \rho$ coincides with the
entanglement of formation. It is given by the von Neumann entropy of
the subsystem characterized by the spin $j_1$:
\begin{equation}
 D(\rho) =  E(\rho) = S(\rho_{j_1})
\end{equation}
where $\rho_{j_1} = {\rm Tr}_{j_2} (\rho)$ is the reduced density matrix of
the first subsystem obtained by tracing out the spin $j_2$ . Thus,
the quantum discord  writes as
\begin{equation}
D(\rho) = - \lambda_+ \log_2 \lambda_+ - \lambda_- \log_2 \lambda_-
\end{equation}
in term of the eigenvalues of the reduced density matrix
$\rho_{j_1}$ given by
\begin{equation}\label{lambda}
\lambda_{\pm}= \frac{1}{2}\bigg( 1 \pm \sqrt{1 - {\cal C}^2} \bigg).
\end{equation}
In Eq.(\ref{lambda}), ${\cal C}$ is the concurrence between the two
subsystems given by
\begin{equation}
{\cal C} = \frac{\sqrt{1-p^{4j_1}}\sqrt{1-p^{4j_2}}}{1+p^{2j}\cos
m\pi}\label{concurence1}
\end{equation}
that is simply obtained by using the qubit mapping (\ref{base0}). It
follows that
 the entanglement of formation writes
\begin{equation}
E_{j_1,j_2} \equiv   E(\rho) =H \bigg(\frac{1}{2} +
\frac{1}{2}\frac{ p^{2j_1} + p^{2j_2}\cos m\pi }{1 + p^{2j}\cos
m\pi}\bigg),\label{qdpure}
\end{equation}
where $H$ stands for the binary entropy defined above. Notice that
the entanglement of formation satisfies the symmetry relation
\begin{equation}\label{sym-E}
E_{j_1,j_2} = E_{j_2,j_1}
\end{equation}
as expected. For $p \rightarrow 0$, the state (\ref{cs-2q}) reduces
to a bipartite state of ${\rm GHZ}$ type which is maximally
entangled $({\cal C} = 1)$ and the entanglement of formation is $
E(\rho)= 1.$ The limiting case  $p \rightarrow 1$ is slightly
different. In fact, we have $ E(\rho)= 0$ for $m$ even  (i.e.
symmetric pure states). The odd spin coherent states (i.e. $m$ odd)
become of $W$ type when $p \rightarrow 1$ and the bipartite
concurrence writes
$${\cal C} = 2~
\frac{\sqrt{j_1j_2}}{j_1+j_2}.$$ It follows that the corresponding
pairwise quantum entanglement takes the form
$$  E(\rho) =  D(\rho)=
H\bigg(\frac{1}{2} + \frac{1}{2}\frac{j_1-j_2}{j_1+j_2} \bigg).$$
The entanglement of formation in $W$ states is maximal when $j_1
= j_2$ ($  E(\rho)= 1$). In other hand, in a splitting scheme such
as $j_2 \ll j_1$ or $j_1 \ll j_2$, the states of $W$ type are unentangled
($ E(\rho)= 0$).

\subsection{Illustration}
To exemplify the above results, we consider the even and odd
coherent states associated with the spin $j=2$. The three possible
bipartite  splitting schemes are
$$ (j_1 = \frac {3}{2} , j_2 = \frac {1}{2}) \quad (j_1 = 1 , j_2 = 1)\quad (j_1 = \frac {1}{2} , j_2 = \frac {3}{2})$$
Using the equation (\ref{qdpure}) and  the relation (\ref{sym-E}),
one gets
\begin{equation}\label{E3/21/2}
E_{\frac {3}{2} ,  \frac {1}{2}}= E_{\frac {1}{2} ,  \frac {3}{2}} =
H \bigg(\frac{1}{2} + \frac{p}{2}\frac{ p^{2} + \cos m\pi }{1 +
p^{4}\cos m\pi}\bigg)
\end{equation}
 and
\begin{equation}\label{E11}
E_{ 1 ,  1} = H \bigg(\frac{1}{2} + \frac{p^{2}}{2}\frac{1  + \cos
m\pi }{1 + p^{4}\cos m\pi}\bigg)
\end{equation}
The behavior of the entanglement of formation $E_{\frac {3}{2} ,
\frac {1}{2}}$ and $E_{ 1 ,  1}$ versus the overlap $p$ is plotted
in the figures 1 and 2 corresponding respectively to even $(m = 0)$
and odd $(m=1)$ spin  coherent states. As seen from the figures, in
both cases the entanglement of formation in the splitting scheme $2
\longrightarrow (1,1)$ is greater than one existing between the spin
subsystems arising from the decomposition $2 \longrightarrow (\frac
{3}{2} ,  \frac {1}{2})$ for any value of $p$. In general, for a
given spin $j$, the maximal value of entanglement of formation
$E_{j_1,j_2}$ is reached in the bipartition where $j_1=j_2=
\frac{j}{2}$. In figure 2, for odd spin coherent states, we have
$E_{ 1 , 1} = 1$ as it can be verified from the expression
(\ref{E11}).
\begin{center}
\includegraphics[width=4in]{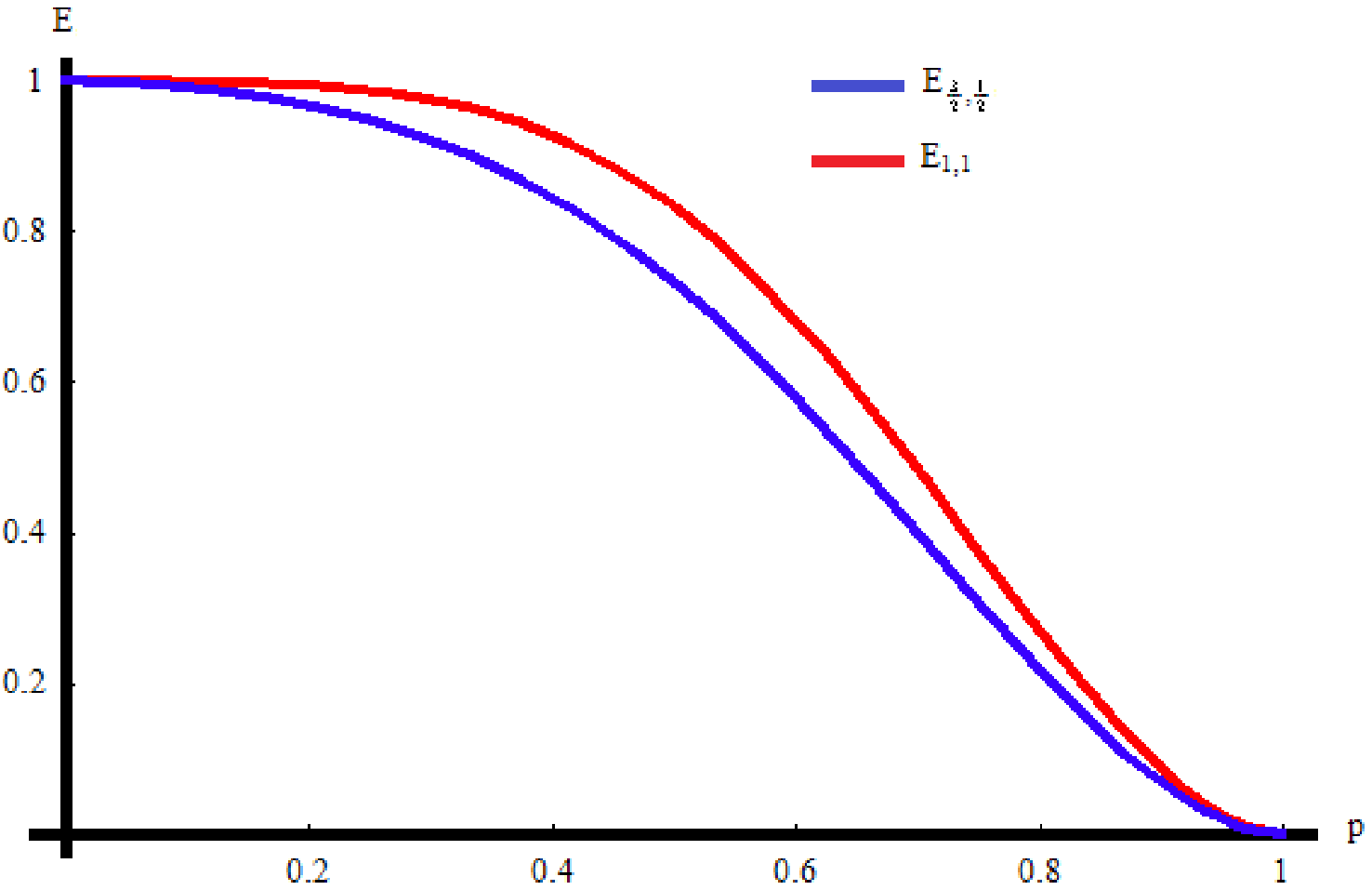}\\
FIG. 1:  {\sf The pairwise entanglement of formation $E =
E_{j_1,j_2}$ versus the overlap $p$ for $(j_1 = \frac{3}{2}, j_2 =
\frac{1}{2})$  and $(j_1 = 1, j_2 = 1)$ with $m = 0$ .}
\end{center}
\begin{center}
\includegraphics[width=4in]{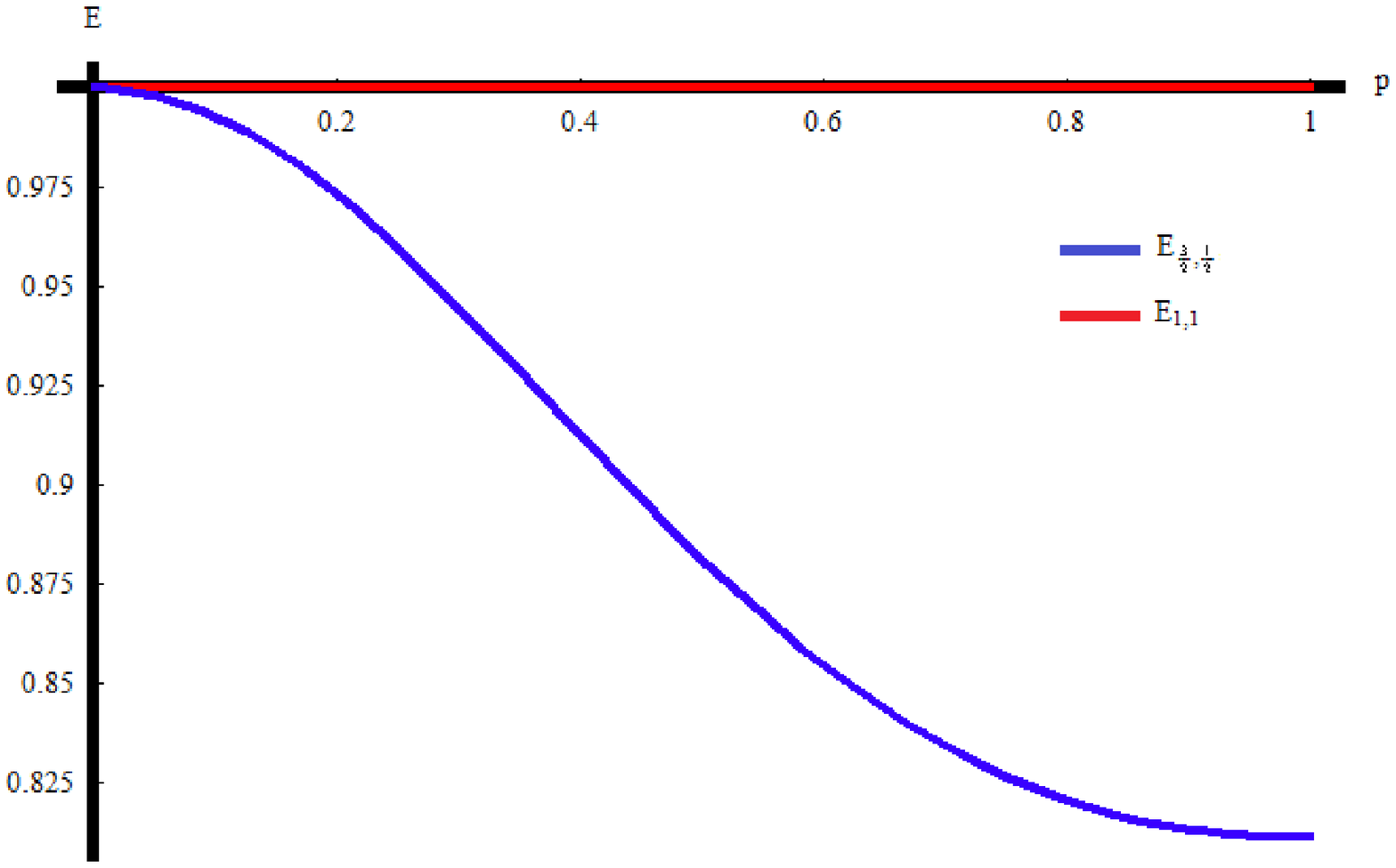}\\
FIG. 2: {\sf The pairwise entanglement of formation $E =
E_{j_1,j_2}$ versus the overlap $p$ for $(j_1 = \frac{3}{2}, j_2 =
\frac{1}{2})$ and $(j_1 = 1, j_2 = 1)$ with $m = 1$ .}
\end{center}

\section{ Three modes splitting and qubit mapping}
Analogously to the bipartite case, we consider in this section the
tripartite splitting of even and odd spin coherent states
(\ref{ncs}).  The entire system decays into three subsystems, one
subsystem describing a particle of spin~$j_1$, the second refers to
a particle of spin~$j_2$ and the remaining particle is of spin $j_3
= j - j_1-j_2$. In this  scheme, the state $\vert j, \eta , m
\rangle$ writes as
\begin{equation}\label{partition3}
 \vert j, \eta , m
\rangle = {\cal N}_m (\vert j_1 , \eta  \rangle \otimes \vert j_2 ,
\eta \rangle \otimes \vert j_3 , \eta \rangle + e^{im\pi } \vert j_1
, -\eta \rangle  \otimes \vert  j_2 , -\eta \rangle \otimes \vert
j_3 , -\eta \rangle).
\end{equation}
To evaluate the bipartite quantum correlations present coherent
states decomposed as in (\ref{partition3}), two different
bi-partitions are considered. The first one yields pure bipartite
states and the second one involves mixed two-qubit states.

\subsection{Bipartite pure  states}
The pure bi-partitions of the state (\ref{partition3}) can be
introduced in three different ways. In the first one, the state $
\vert j, \eta , m \rangle$ is written as
\begin{equation}\label{partition31}
 \vert j, \eta , m
\rangle_{j_1 \vert j-j_1} = {\cal N}_m (\vert j_1 , \eta \rangle
\otimes \vert j-j_1 , \eta \rangle  + e^{im\pi } \vert  j_1 , - \eta
\rangle \otimes \vert j- j_1 , -\eta \rangle).
\end{equation}
Similarly, the state (\ref{partition3}) can be also partitioned as
\begin{equation}\label{partition32}
 \vert j, \eta , m
\rangle_{j_2 \vert j-j_2} = {\cal N}_m (\vert j_2 , \eta \rangle
\otimes \vert j-j_2 , \eta \rangle  + e^{im\pi } \vert  j_2 , - \eta
\rangle \otimes \vert j- j_2 , -\eta \rangle).
\end{equation}
The third bipartition is given by
\begin{equation}\label{partition33}
 \vert j, \eta , m
\rangle_{j_3 \vert j-j_3} = {\cal N}_m (\vert j_3 , \eta \rangle
\otimes \vert j-j_3 , \eta \rangle  + e^{im\pi } \vert  j_3 , - \eta
\rangle \otimes \vert j- j_3 , -\eta \rangle).
\end{equation}
For each bipartition, the  state $\vert j, \eta , m \rangle$ can be
converted into a state of two logical qubits. This is achieved by
introducing, for the first subsystem, the orthogonal basis $\{ \vert
0 \rangle_l , \vert 1 \rangle_l\}$, with $ l = j_1, j_2$ or $j_3$,
defined as
\begin{equation}\label{base1}
\vert 0 \rangle_l = \frac{ \vert l , \eta  \rangle +  \vert l ,
-\eta \rangle}{\sqrt{2(1 + p^{2l})}}
   \qquad \vert 1 \rangle_l = \frac{\vert l , \eta \rangle -  \vert l , -\eta
\rangle}{{\sqrt{2(1- p^{2l})}}},
\end{equation}
and, for the second subsystem, the orthogonal basis $\{ \vert  0
\rangle_{j-l} , \vert  1 \rangle_{j-l}\}$ given by
\begin{equation}\label{base2}
\vert 0 \rangle_{j-l} = \frac{ \vert  j-l , \eta  \rangle + \vert
j-l , -\eta \rangle}{\sqrt{2(1 + p^{2(j-l)})}}
   \qquad \vert 1 \rangle_{j-l} = \frac{\vert  j-l , \eta \rangle -  \vert  j-l , -\eta
\rangle}{{\sqrt{2(1- p^{2(j-l)})}}}.
\end{equation}
Reporting the equations (\ref{base1}) and (\ref{base2}) in
(\ref{partition31}), (\ref{partition32}) and (\ref{partition33}),
one has the expression of the pure state $ \vert j, \eta , m
\rangle_{l \vert j- l}$ in the basis $\{ \vert 0 \rangle_{l} \otimes
\vert 0 \rangle_{j-l} ,
 \vert 0 \rangle_{l} \otimes \vert 1 \rangle_{j-l} , \vert  1 \rangle_{l}
 \otimes \vert 0 \rangle_{j-l} , \vert  1 \rangle_{l} \otimes \vert  1
 \rangle_{j-l}\}$. It is given by
\begin{equation}
 \vert j, \eta , m
\rangle_{l \vert j-l} = \sum_{\alpha= 0,1} \sum_{\beta= 0,1}
C_{\alpha,\beta} \vert \alpha \rangle_l \otimes \vert \beta
\rangle_{j-l}\label{mapping1}
\end{equation}
where the coefficients $C_{\alpha,\beta}$ are
$$ C_{0,0} = {\cal N}_m(1 + e^{im\pi}) a_{l}a_{j-l}  , \qquad  C_{0,1} =  {\cal N}_m (1 -e^{im\pi}) a_{l}b_{j-l} $$
$$ C_{1,0} = {\cal N}_m (1 - e^{im\pi}) a_{j-l}b_{l}  , \qquad  C_{1,1} =  {\cal N}_m (1 + e^{im\pi}) b_{l}b_{j-l}. $$
in terms of the quantities
$$ a_k =\sqrt{\frac{1+p^{2k}}{2}} , \qquad b_k = \sqrt{\frac{1-p^{2k}}{2}} \qquad {\rm for} ~ k = l, j-l$$
involving the overlap $p$ (\ref{overlap}) which is related to the
non-orthogonality of two spin coherent states of equal amplitude and
opposite phase.

\subsection{Bipartite mixed states}
The second class of bipartite density matrices can be realized from
the state (\ref{partition3}) by considering the  reduced density
matrices $\rho_{l_1 l_2}$ that are obtained by tracing out the
degrees of freedom of the third subsystem. There are three different
density matrices  $ \rho_{j_1 j_2}$ , $\rho_{j_2 j_3}$ and $
\rho_{j_1 j_3}$. Explicitly,  they are given by
\begin{eqnarray}
\rho_{l_1l_2} &=&\text{Tr}_{l_3}(\vert j, \eta , m\rangle \langle
j, \eta , m \vert)  \nonumber \\
&=&{\cal N}_m^2(\vert \eta , \eta )( \eta , \eta \vert +\vert - \eta
, - \eta )( - \eta , - \eta | + e^{i m \pi } q |- \eta , - \eta )(
\eta , \eta  \vert +e^{-i m \pi }q\vert \eta , \eta )( - \eta, -
\eta \vert ) \label{rho12}
\end{eqnarray}
with $q \equiv p^{2(j-l_1-l_2)} = p^{2l_3}$ and
$$\vert \pm \eta , \pm \eta ) = \vert  l_1, \pm \eta  \rangle \otimes \vert  l_2, \pm \eta \rangle. $$
It is interesting to note that the density matrix $\rho_{l_1l_2}$ is a
two-rank  operator. Indeed, it  rewrites as
\begin{eqnarray}
\rho_{l_1l_2} &=& \frac{1}{2} (1 + q)~ \frac{{\cal N}_m^2}{{\cal
N}_+^2}~ \vert \phi_+ \rangle \langle \phi_+ \vert +  \frac{1}{2} (1
- q)~ \frac{{\cal N}_m^2}{{\cal N}_-^2} ~\vert \phi_- \rangle,
\langle \phi_- \vert
\end{eqnarray}
where
$$\vert \phi_{\pm} \rangle = {\cal N}_{\pm} ( \vert  l_1, \eta  \rangle \otimes \vert  l_2,\eta \rangle \pm e^{im\pi}
\vert  l_1, -\eta  \rangle \otimes \vert  l_2, -\eta \rangle)$$ and
$${\cal N}_{\pm}^2 = 2 \pm 2  p^{2(l_1+l_2)} \cos m\pi.$$
In this case, the density matrix $\rho_{l_1l_2}$ can be also
converted into a two-qubit system by an appropriate qubit mapping.
For this, we introduce an orthogonal pair $\{\vert 0\rangle_l ,\vert
1\rangle_l \}$ as
\begin{equation}\label{base}
\vert 0 \rangle_l = \frac{ \vert l, \eta  \rangle +  \vert l,  -\eta
\rangle}{\sqrt{2(1 + p^{2l})}}
   \qquad \vert 1 \rangle_l = \frac{\vert l , \eta \rangle -  \vert  l, -
   \eta
\rangle}{{\sqrt{2(1- p^{2l})}}}.
\end{equation}
where $l = l_1$ for the first subsystem and $l = l_2$ for the
second. Substituting the equation (\ref{base}) into (\ref{rho12}),
we obtain the density matrix
\begin{equation}
\rho_{l_1l_2} = {\cal N}^2 \left( \begin{smallmatrix}
2a_1^2a_2^2(1+q\cos m\pi)& 0
    & 0& 2a_1b_1a_2b_2(1+q\cos m\pi
)\\
0  & 2a_1^2b_2^2(1-q\cos m\pi ) & 2a_1b_1a_2b_2(1-q\cos m\pi
) & 0 \\
0  & 2a_1b_1a_2b_2(1-q\cos m\pi ) & 2a_2^2b_1^2(1-q\cos m\pi
) & 0 \\
2a_1b_1a_2b_2(1+q\cos m\pi ) & 0 & 0 & 2b_1^2b_2^2(1+q\cos m\pi)
\end{smallmatrix}
\right) \label{rho12-matrix}
\end{equation}
in the basis $\{\vert 0_{l_1}, 0_{l_2} \rangle ,\vert 0_{l_1},
1_{l_2}\rangle ,\vert 1_{l_1},0_{l_2}\rangle ,
    \vert  1_{l_1}, 1_{l_2}\rangle \}$ where the quantities $a_1, b_1, a_2,
    b_2$ are defined by

$$ a_i =\sqrt{\frac{1+p^{2l_i}}{2}} , \qquad b_i = \sqrt{\frac{1-p^{2l_i}}{2}} \qquad {\rm for} ~ i = 1, 2$$

\section{ Quantum entanglement in the three splitting scheme}

\subsection{ Entanglement of formation}
In the pure bipartite splitting scheme, the concurrence is given by
\begin{equation}
{\cal C}(\rho_{k_1\vert k_2k_3}) =
=\frac{\sqrt{1-p^{4k_1}}\sqrt{1-p^{4(j-k_1)}}}{1+p^{2j}\cos m\pi}
\end{equation}
where the triplet $(k_1, k_2, k_3)$ stands  for $(j_1, j_2, j_3)$ ,
$(j_2, j_1, j_3)$  and $(j_3, j_1, j_2)$ corresponding respectively
to the states (\ref{partition31}), (\ref{partition32}) and
(\ref{partition33}). Subsequently, the entanglement of formation
writes
\begin{equation}\label{E3pure}
E(\rho_{k_1\vert k_2k_3}) = H \bigg(\frac{1}{2} + \frac{1}{2}\frac{
p^{2k_1} + p^{2(j-k_1)}\cos m\pi }{1 + p^{2j}\cos m\pi}\bigg).
\end{equation}
For mixed bipartite states belonging to the second bi-partitioning
class (\ref{rho12}), the concurrence  is given by
\begin{eqnarray}
{\cal C}(\rho_{l_1l_2}) = p^{2(j-l_1-l_2)}
~\frac{\sqrt{(1-p^{4l_1})(1-
 p^{4l_2})}}{1 + p^{2j} \cos m\pi}
\end{eqnarray}
where the reduced density matrix $\rho_{l_1l_2}$ stands for
$\rho_{j_1j_2}$, $\rho_{j_2j_3}$ and $\rho_{j_1j_3}$. The
entanglement of formation writes
\begin{equation}\label{E3mixte}
E(\rho_{l_1l_2}) = H\bigg(\frac{1}{2} + \frac{1}{2} \sqrt{1 -
 ~\frac{p^{4(j-l_1-l_2)}(1-p^{4l_1})(1-
 p^{4l_2})}{(1 + p^{2j} \cos m\pi)^2}}\bigg).
\end{equation}

\subsection{Multipartite entanglement of formation}
When the bipartite quantum correlations are quantified by the
entanglement of formation, the definition (\ref{Qtotal}) gives
\begin{eqnarray}
E(j_1,j_2,j_3) = \frac{1}{6}(E(\rho_{j_1j_2}) + E(\rho_{j_1j_3}) +
E(\rho_{j_2j_3}) + E(\rho_{j_1\vert j_2j_3})+ E(\rho_{j_2\vert
j_1j_3}) + E(\rho_{j_3\vert j_1j_2}))
\end{eqnarray}
Using the results (\ref{E3pure}) and (\ref{E3mixte}), the total
amount of quantum entanglement is explicitly  given by
\begin{eqnarray}\label{Etotal}
E(j_1,j_2,j_3) &=& \frac{1}{6} \bigg[H \bigg(\frac{1}{2} +
\frac{1}{2}\frac{ p^{2j_1} + p^{2(j_2+j_3)}\cos m\pi }{1 +
p^{2j}\cos m\pi}\bigg)+H \bigg(\frac{1}{2} + \frac{1}{2}\sqrt{ 1 -
\frac{p^{4j_1} (1 - p^{4j_2})(1 - p^{4j_3}) }{(1 + p^{2j}\cos
m\pi)^2}}\bigg) \nonumber\\
&+& H \bigg(\frac{1}{2} + \frac{1}{2}\frac{ p^{2j_2} +
p^{2(j_1+j_3)}\cos m\pi }{1 + p^{2j}\cos m\pi}\bigg)+ H
\bigg(\frac{1}{2} + \frac{1}{2}\sqrt{ 1 - \frac{p^{4j_2} (1 -
p^{4j_1})(1 - p^{4j_3}) }{(1 + p^{2j}\cos m\pi)^2}}\bigg)\nonumber\\
&+& H \bigg(\frac{1}{2} + \frac{1}{2}\frac{ p^{2j_3} +
p^{2(j_1+j_2)}\cos m\pi }{1 + p^{2j}\cos m\pi}\bigg)+H
\bigg(\frac{1}{2} + \frac{1}{2}\sqrt{ 1 - \frac{p^{4j_3} (1 -
p^{4j_1})(1 - p^{4j_2}) }{(1 + p^{2j}\cos m\pi)^2}}\bigg) \bigg]~~~
\end{eqnarray}
which is completely symmetric in $j_1$, $j_2$ and $j_3$. This
quantity will be compared with the sum of pairwise quantum discord
of all possible bi-partitions of the state (\ref{partition3})  and
its behavior in terms of the overlap $p$ in some particular cases is
examined in Section 7.

\subsection{Monogamy of entanglement of formation}

The entanglement shared by more than two parties constitutes a
subtle issue in investigating  multipartite correlations. Thus,
considering the limitations of sharing entanglement in the
orthogonal case, we study the monogamy of entanglement of formation
in tripartite spin coherent states. In this respect, we analyze the
situations where  the following inequality
$$E(\rho_{l_1l_2}) + E(\rho_{l_1l_3}) \leq E(\rho_{l_1\vert l_2l_3})$$
is satisfied or violated. The notations are as above. Clearly, to
decide if the entanglement of formation is monogamous or not in spin
coherent states, we shall treat some particular cases. We first
consider the splitting $j_1 = j_2 = j_3 = \frac{1}{2}$ which arises
from the decomposition of even and odd coherent states associated
with the spin $j = \frac{3}{2}$. The behavior of the entanglement of
formation difference :
$$\Delta E = E(\rho_{j_1\vert j_2j_3}) - E(\rho_{j_1j_2}) - E(\rho_{j_1j_2}),$$
for even and odd spin coherent states, are reported in the figure 3.
They show that the entanglement of formation satisfies always the
monogamy relation in the even case $(m = 0)$ but ceases to be
monogamous in the odd case $(m = 1)$ when the overlap $p$ is greater
than 0.8. This indicates also that the monogamy relation is violated
in three qubit states of $W$ type obtained in the limiting case
$p \longrightarrow 1$. Similarly, we also considered the two
tripartite splitting $(j_1 = \frac{1}{2},j_2 = \frac{1}{2},j_3 = 1)$
and $(j_1 = 1,j_2 = \frac{1}{2},j_3 = \frac{1}{2})$ which can
originate from the splitting of the spin $j=2$. The figures 4
reveals that the monogamy relation is satisfied for even spin
coherent states ($m = 0$).  However, for odd spin coherent states
($m = 1$), the entanglement of formation does not follow the
monogamy as $p$ approaches the unity (see figure 4). This agrees
with the result of figure 3 and confirms that in a $W$ state
comprising three qubits, the monogamy of entanglement of formation
is violated.
\begin{center}
\includegraphics[width=4in]{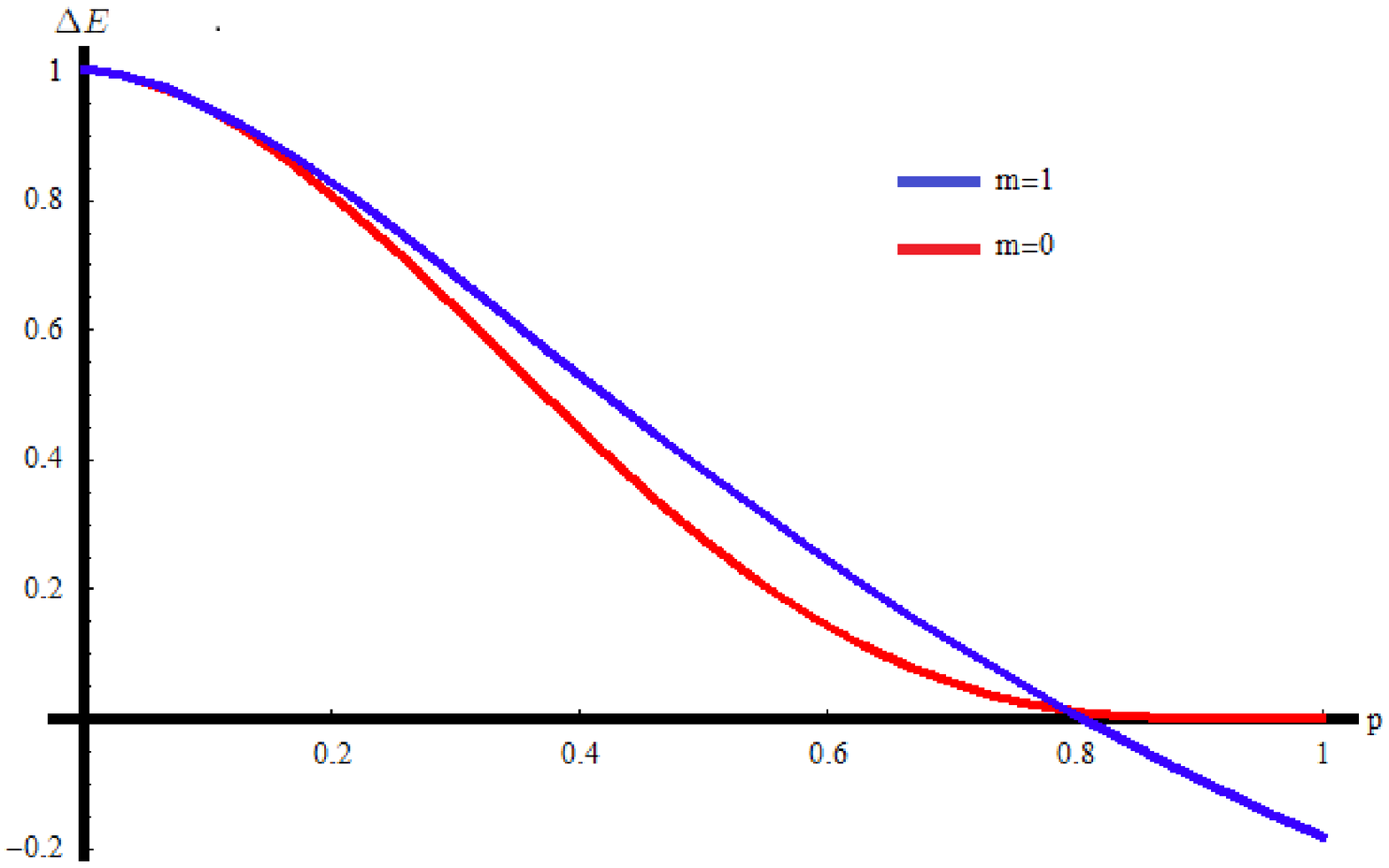}\\
FIG. 3:  {\sf The function $\Delta E $ versus the overlap $p$ when
$j_1=j_2=j_3=\frac{1}{2}$ for $m=0$ and $m=1$.}
\end{center}
\begin{center}
\includegraphics[width=4in]{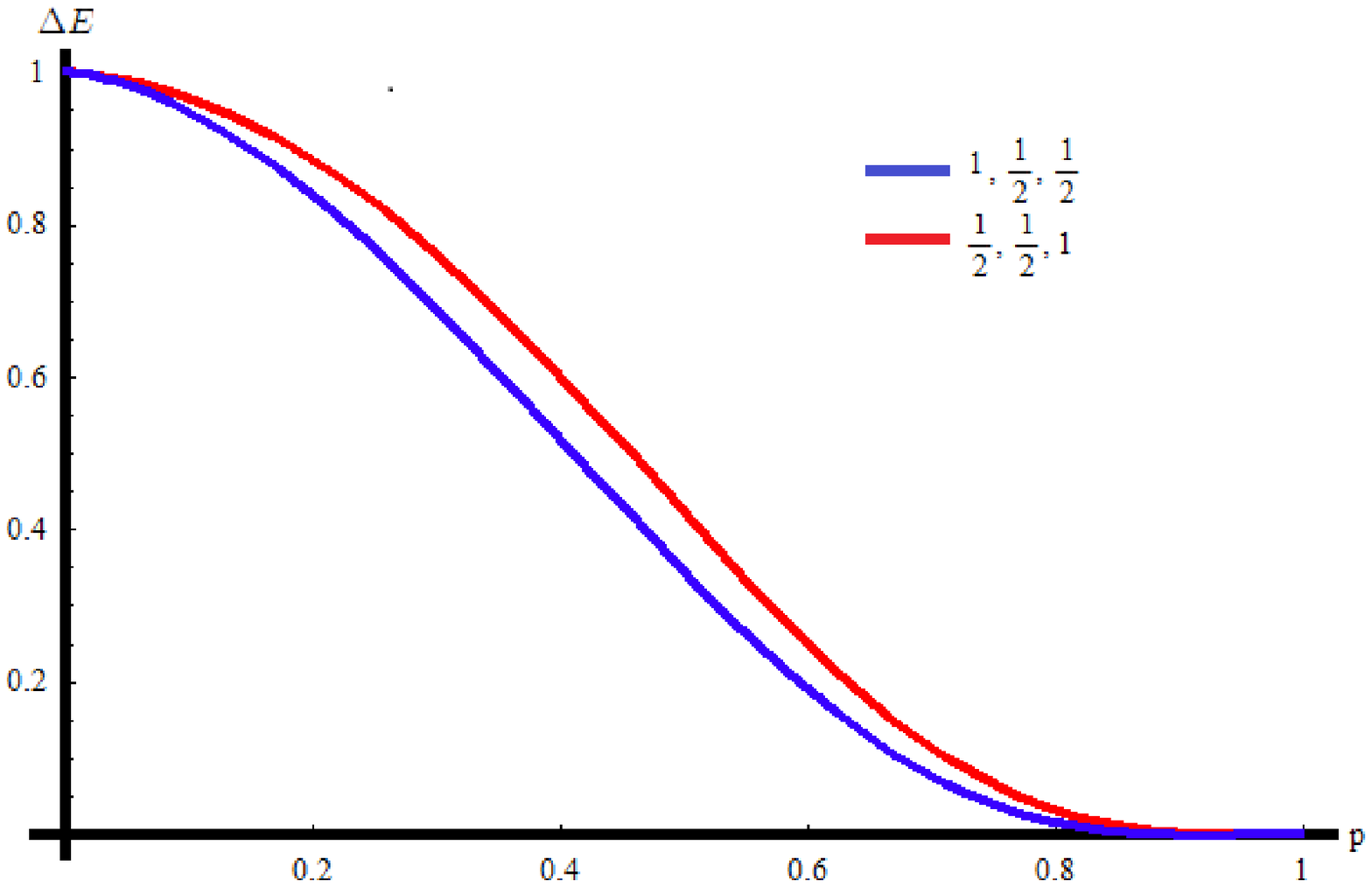}\\
FIG. 4:  {\sf The function $\Delta E $ versus the overlap $p$ when
$(j_1= \frac{1}{2}, j_2= \frac{1}{2}, j_3 = 1)$ and $(j_1= 1, j_2=
\frac{1}{2}, j_3 = \frac{1}{2})$   for $m=0$.}
\end{center}
\begin{center}
\includegraphics[width=4in]{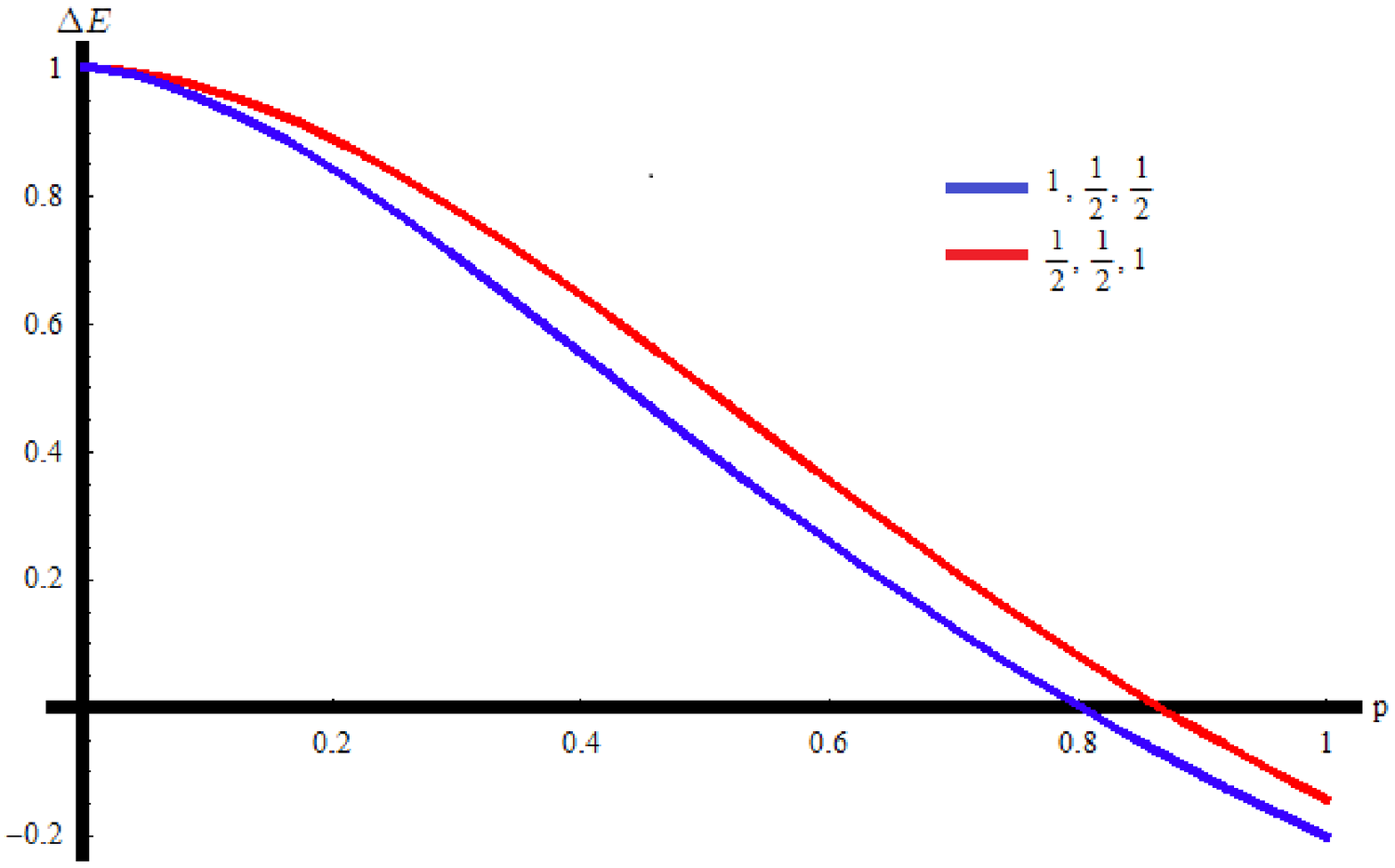}\\
FIG. 5:   {\sf The function $\Delta E $ versus the overlap $p$ when
$(j_1= \frac{1}{2}, j_2= \frac{1}{2}, j_3 = 1)$ and $(j_1= 1, j_2=
\frac{1}{2}, j_3 = \frac{1}{2})$   for $m=1$.}
\end{center}

\section{ Quantum discord in the three splitting scheme}

\subsection{ Quantum discord }

In  the pure bipartite splitting scheme defined by
(\ref{partition31}), (\ref{partition32}) and (\ref{partition33}),
the quantum discord and entanglement of formation as measure of
bipartite quantum correlations are identical and we have
\begin{equation}
D(\rho_{j_1\vert j_2j_3}) = E(\rho_{j_1\vert j_2j_3}) \qquad
D(\rho_{j_2\vert j_1j_3}) = E(\rho_{j_2\vert j_1j_3}) \qquad
D(\rho_{j_3\vert j_1j_2}) = E(\rho_{j_1\vert j_1j_2})
\end{equation}
where the entanglement of formation is given by (\ref{E3pure})
modulo some obvious substitutions.

\noindent To get the explicit expressions of quantum discord in
 bipartite mixed states $\rho_{l_1l_2}$ of the form  (\ref{rho12-matrix}), we  evaluate the mutual
information entropy and the minimum of conditional entropy according
to the general algorithm discussed in Section 2. We first calculate
the mutual information. The non vanishing eigenvalues of the density
matrix $\rho_{l_1l_2}$ are
\begin{equation}
\lambda_{\pm} = \frac{1}{2} \frac{(1\pm p^{2(j- l_1- l_2)})(1 \pm
p^{2(l_1+l_2)}\cos(m\pi))}{1 + p^{2j}\cos(m\pi)},\label{lambda+-}
\end{equation}
and the joint entropy is
\begin{equation}\label{entropy12}
S(\rho_{l_1l_2}) = h(\lambda_+) + h(\lambda_-) = H(\lambda_+).
\end{equation} The eigenvalues of the marginal $\rho_{l_1} = {\rm
Tr}_{l_2} \rho_{l_1l_2}$ are
$$\lambda_{1,\pm} = \frac{1}{2} \frac{(1\pm p^{2(j-l_1)})(1 \pm p^{2l_1}\cos(m\pi))}{1 + p^{2j}\cos(m\pi)},$$
and the marginal entropy reads
\begin{equation}\label{entropy1}
S(\rho_{l_1}) = h(\lambda_{1,+}) + h(\lambda_{1,-}) =
H(\lambda_{1,+}).
\end{equation}
The eigenvalues of the marginal $\rho_{l_2} = {\rm Tr}_{l_1}
\rho_{l_1l_2}$ are
$$\lambda_{2,\pm} = \frac{1}{2} \frac{(1\pm p^{2(j-l_2)})(1 \pm p^{2l_2}\cos(m\pi))}{1 + p^{2j}\cos(m\pi)},$$
and the corresponding entropy is given by
\begin{equation}\label{entropy2}
S(\rho_{l_2}) = h(\lambda_{2,+}) + h(\lambda_{2,-}) =
H(\lambda_{2,+}).
\end{equation}
It follows that  the mutual information defined by (\ref{def: mutual
information}) takes the form
\begin{equation}
    I(\rho_{l_1l_2})= H(\lambda_{1,+}) + H(\lambda_{2,+}) - H(\lambda_+) .
\end{equation}
The second important step in deriving pairwise quantum discord
requires the explicit calculation of the minimal amount of the
conditional entropy (\ref{condit-entropy}).  According the general
discussion presented in the second section, it is necessary to
purify the density matrix $\rho_{l_1l_2}$ and determine the entanglement of
formation of its complement.  This  algorithm can be achieved as
follows. The matrix $\rho_{l_1l_2}$ is a two-qubit state and
subsequently decomposes as
\begin{eqnarray}
\rho_{l_1l_2} = \lambda_+ \vert \phi_+ \rangle \langle \phi_+ \vert
+ \lambda_- \vert \phi_- \rangle \langle \phi_- \vert
\end{eqnarray}
where the eigenvalues $\lambda_+$ and $\lambda_-$ are given by
(\ref{lambda+-}) and the corresponding eigenstates $\vert \phi_+
\rangle$ and $\vert \phi_- \rangle$ write as
\begin{eqnarray}
\vert \phi_+ \rangle =
\frac{\sqrt{(1+p^{l_1})(1+p^{l_2})}}{\sqrt{2(1+p^{l_1+l_2})}} \vert
 0_{l_1} ,  0_{l_2} \rangle +
\frac{\sqrt{(1-p^{l_1})(1-p^{l_2})}}{\sqrt{2(1+p^{l_1+l_2})}} \vert
 1_{l_1},  1_{l_2} \rangle
\end{eqnarray}

\begin{eqnarray}
\vert \phi_- \rangle =
\frac{\sqrt{(1+p^{l_1})(1-p^{l_2})}}{\sqrt{2(1(-p^{l_1+l_2})}} \vert
0_{l_1} ,  1_{l_2} \rangle +
\frac{\sqrt{(1-p^{l_1})(1+p^{l_2})}}{\sqrt{2(1-p^{l_1+l_2})}} \vert
 1_{l_1} ,  0_{l_2} \rangle
\end{eqnarray}
in the basis (\ref{base}). Attaching a qubit $3$ to the two-qubit
system $(12)\equiv(l_1l_2)$, we write the purification of
$\rho_{l_1l_2}$ as
\begin{eqnarray}
\vert \phi \rangle = \sqrt{\lambda_+} \vert \phi_+ \rangle \otimes
\vert {\bf 0}  \rangle +  \sqrt{\lambda_-} \vert \phi_- \rangle
\otimes \vert {\bf 1} \rangle
\end{eqnarray}
such that the whole system $(123)$ is described by the pure density
matrix $\rho_{l_1l_23} = \vert \phi \rangle \langle \phi \vert $.
Using the Koashi-Winter relation (\ref{stild-min}), we have
\begin{equation}\label{smin}
\widetilde{S}_{\rm min} = E(\rho_{23}) = H(\frac{1}{2} + \frac{1}{2}
\sqrt{1 - \vert {\cal C}(\rho_{23})\vert^2})
\end{equation}
where the concurrence of the density matrix $\rho_{23}\equiv \rho_{l_23}$
is
$$\vert {\cal C}(\rho_{l_23})\vert^2 =  \frac{p^{4l_1}(1 - p^{4l_2})(1 - p^{4(j-l_1-l_2)})}{(1+p^{2j}\cos m\pi)^2}.$$
It follows that the quantum discord is then given by
\begin{eqnarray}
D(\rho_{l_1l_2}) = S(\rho_{l_1}) - S(\rho_{l_1l_2}) +
E(\rho_{l_23}).
\end{eqnarray}
Using the equations (\ref{entropy12}), (\ref{entropy1}) and
(\ref{smin}), it rewrites explicitly as
\begin{eqnarray}\label{qd-right}
D^{\rightarrow}(\rho_{l_1l_2}) &=& H\bigg( \frac{1}{2} \frac{(1 +
p^{2l_1})(1 + p^{2(j-l_1)}\cos(m\pi))}{1 +
p^{2j}\cos(m\pi)}\bigg)\nonumber \\ &-& H\bigg(\frac{1}{2} \frac{(1+
p^{2(j- l_1- l_2)})(1 +
p^{2(l_1+l_2)}\cos(m\pi))}{1 + p^{2j}\cos(m\pi)}\bigg)\\
&+& H\bigg(\frac{1}{2} + \frac{1}{2} \sqrt{1 -  \frac{p^{4l_1}(1 -
p^{4l_2})(1 - p^{4(j-l_1-l_2)})}{(1+p^{2j}\cos
m\pi)^2}}\bigg)\nonumber
\end{eqnarray}
where the pair $(l_1,l_2)$ stands for $(j_1,j_2)$, $(j_1,j_3)$ and
$(j_2,j_3)$. Similarly, the measure of quantum discord obtained by
measuring the second qubit $B\equiv l_2$ is
\begin{eqnarray}\label{qd-left}
D^{\leftarrow}(\rho_{l_1l_2}) &=& H\bigg( \frac{1}{2} \frac{(1 +
p^{2l_2})(1 + p^{2(j-l_2)}\cos(m\pi))}{1 +
p^{2j}\cos(m\pi)}\bigg)\nonumber \\ &-& H\bigg(\frac{1}{2} \frac{(1+
p^{2(j- l_1- l_2)})(1 +
p^{2(l_1+l_2)}\cos(m\pi))}{1 + p^{2j}\cos(m\pi)}\bigg)\\
&+& H\bigg(\frac{1}{2} + \frac{1}{2} \sqrt{1 -  \frac{p^{4l_2}(1 -
p^{4l_1})(1 - p^{4(j-l_1-l_2)})}{(1+p^{2j}\cos
m\pi)^2}}\bigg).\nonumber
\end{eqnarray}
It is interesting to note that
\begin{eqnarray}\label{qd-right-left}
D^{\rightarrow}(\rho_{l_1l_2}) = D^{\leftarrow}(\rho_{l_2l_1}).
\end{eqnarray}
It is clear that for $l_1=l_2$, the quantum discord is symmetric,
i.e. $D^{\rightarrow}(\rho_{ll}) = D^{\leftarrow}(\rho_{ll})$. Using
the equation (\ref{qd-right}), one obtains the  following
conservation relations
\begin{eqnarray}\label{sum-discord}
D^{\rightarrow}(\rho_{j_1j_2}) + D^{\rightarrow}(\rho_{j_3j_2}) =
E_{j_2j_3} + E_{j_2j_1},\nonumber \\
D^{\rightarrow}(\rho_{j_2j_1}) + D^{\rightarrow}(\rho_{j_3j_1}) =
E_{j_1j_3} + E_{j_1j_2}, \\
 D^{\rightarrow}(\rho_{j_1j_3})
+ D^{\rightarrow}(\rho_{j_2j_3}) = E_{j_3j_2} + E_{j_3j_1}.\nonumber
\end{eqnarray}
Similar conservations relations hold  for the measures of quantum
discord given by (\ref{qd-left}). They  can be  easily derived from
the relation (\ref{qd-right-left}). Using the conservation relations
(\ref{sum-discord}), we have
$$D^{\rightarrow}(\rho_{j_1j_2}) + D^{\rightarrow}(\rho_{j_2j_3})+  D^{\rightarrow}(\rho_{j_3j_1}) = E_{j_1j_2} + E_{j_1j_3} + E_{j_2j_3}.$$
This reflects that the sum of pairwise quantum discord for all
bipartite mixed states coincides with the sum of entanglement of
formation. It must be noticed that the conservation relations of type (72)
involving entanglement of formation and quantum discord were first derived
in \cite{Z-H Ma}.

\subsection{ Multipartite quantum correlations}

Based on the asymmetric definition of quantum discord, two
interesting quantities were  defined by Fanchini et al
\cite{Fanchini}. In our context, they write
\begin{eqnarray}\label{delta+}
\Delta^+_{l_1\vert l_2} = \frac{1}{2} \big(
D^{\rightarrow}(\rho_{l_1l_2}) +
D^{\rightarrow}(\rho_{l_2l_1})\big),
\end{eqnarray}
and
\begin{eqnarray}\label{delta-}
\Delta^-_{l_1\vert l_2} = \frac{1}{2} \big(
D^{\rightarrow}(\rho_{l_1l_2}) -
D^{\rightarrow}(\rho_{l_2l_1})\big).
\end{eqnarray}
The sum $\Delta^+_{l_1\vert l_2}$ is the average of locally
inaccessible information when the measurements are performed on the
subsystems $l_1$ and $l_2$. It quantifies the disturbance caused by
any local measurement. The  difference $\Delta^-_{l_1\vert l_2}$ is
the balance of locally inaccessible information and quantifies the
asymmetry between the subsystems in responding to the measurement
disturbance. Using the equation (\ref{qd-right}), it is easy to
verify that the average and the balance of quantum discord satisfy
the following identities
\begin{eqnarray}\label{conser-DE+}
\Delta^+_{j_1\vert j_2} + \Delta^+_{j_1\vert j_3} +
\Delta^+_{j_2\vert j_3} = E_{j_1j_2} + E_{j_1j_3} + E_{j_2j_3},
\end{eqnarray}
and
\begin{eqnarray}\label{conser-DE-}
\Delta^-_{j_1\vert j_2} + \Delta^-_{j_1\vert j_3} +
\Delta^-_{j_2\vert j_3} = 0.
\end{eqnarray}
Using the main definition (\ref{Qtotal}), it is interesting to note
that  the total amount of quantum discord present in the state
(\ref{partition3}) can be simply written in terms of  the average of
locally inaccessible information (\ref{delta+}). Indeed, we have
\begin{eqnarray}
D(j_1,j_2,j_3) = \frac{1}{6} \bigg( \Delta^+_{j_1\vert j_2} +
\Delta^+_{j_1\vert j_3}+ \Delta^+_{j_2\vert j_3} +
\Delta^+_{j_1\vert (j_2j_3)}+\Delta^+_{j_2\vert
(j_1j_3)}+\Delta^+_{j_3\vert (j_1j_2)}\bigg)
\end{eqnarray}
where the quantity $\Delta^+_{k_1\vert (k_2k_3)}$ coincides with the
entanglement of formation $E(\rho_{k_1\vert k_2k_3})$ given by
(\ref{E3pure}). Furthermore, using the conservation relation
(\ref{conser-DE+}), one gets
\begin{eqnarray}\label{Dtotal}
D(j_1,j_2,j_3) &=& E(j_1,j_2,j_3)
\end{eqnarray}
where $E(j_1,j_2,j_3)$ is given by (\ref{Etotal}).
This result coincides with  one obtained in \cite{Z-H Ma}.
It reflects that  the sum of quantum discord present in all possible bi-partitions
is exactly the total amount of bipartite entanglement of formation in the entire system.


Since for a spin-$j$ coherent state there are different tripartite
splitting possibilities denoted here by $(j_1,j_2,j_3)$ such that
$j_1+j_2+j_3 = j$, it is seems natural to compare the total amount
of multipartite correlations in each splitting scheme. As
illustration, we consider the situation where $j = 3$. The
tripartite quantum discord $D(j_1, j_2, j_3)$ (\ref{Dtotal}) is
totally symmetric in $j_1$, $j_2$ and $j_3$. Thus, for $j=3$, three
inequivalent splitting schemes are of special interest. They
correspond to $(j_1 = 1, j_2 = 1, j_3 = 1)$ , $(j_1 = \frac{1}{2},
j_2 = \frac{1}{2} , j_3 = 2)$ and $(j_1 = \frac{1}{2}, j_2 = 1  ,
j_3 = \frac{3}{2})$. In figures 6 and 7, we plot the quantity
$D(j_1,j_2,j_3) $ as
function of the overlap $p$ for each case.\\



\begin{center}
\includegraphics[width=4in]{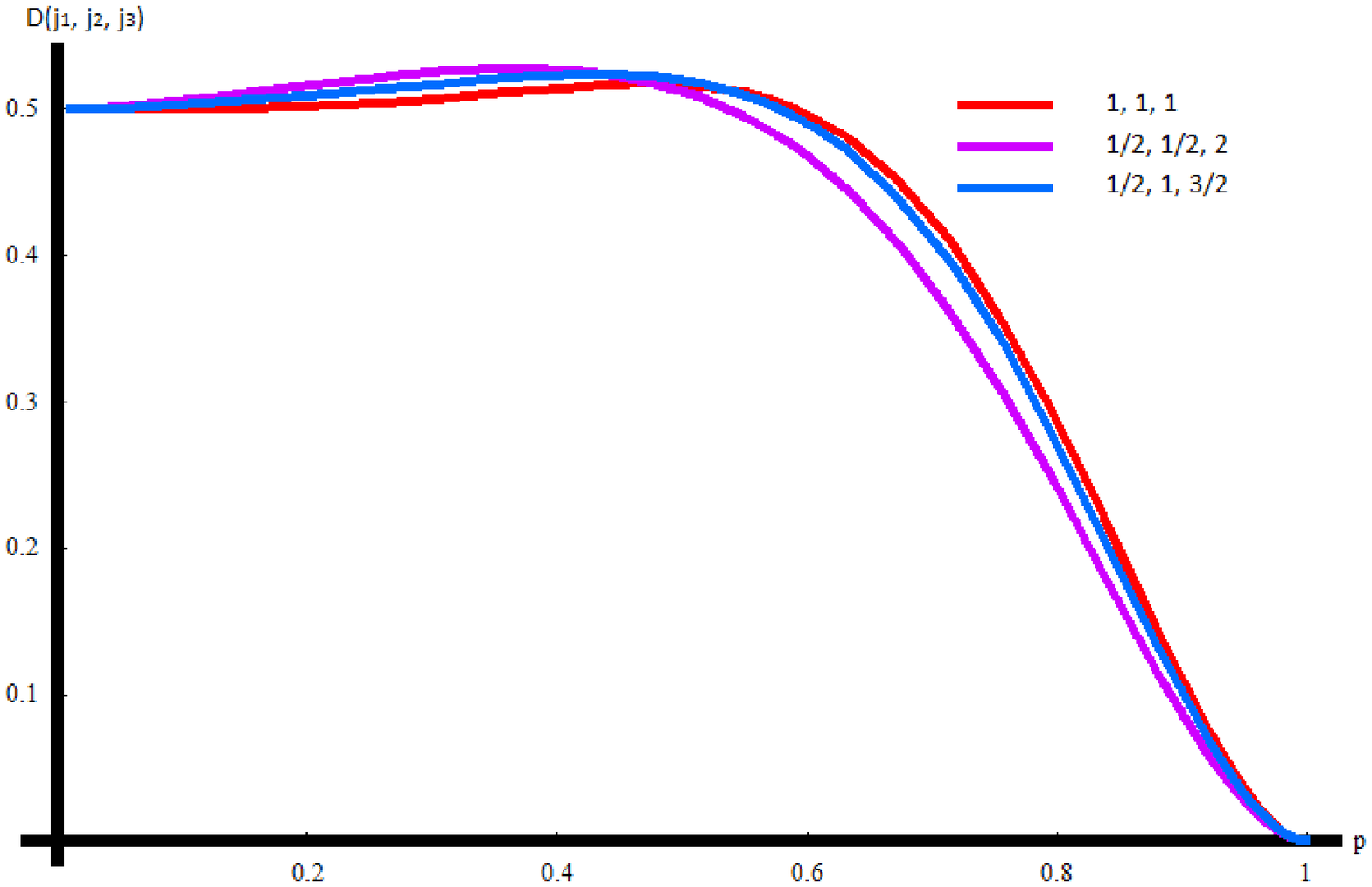}\\
FIG. 6:  {\sf The multipartite quantum correlations for  $j=3$
versus the overlap $p$ for $m=0$.}
\end{center}

\begin{center}
\includegraphics[width=3in]{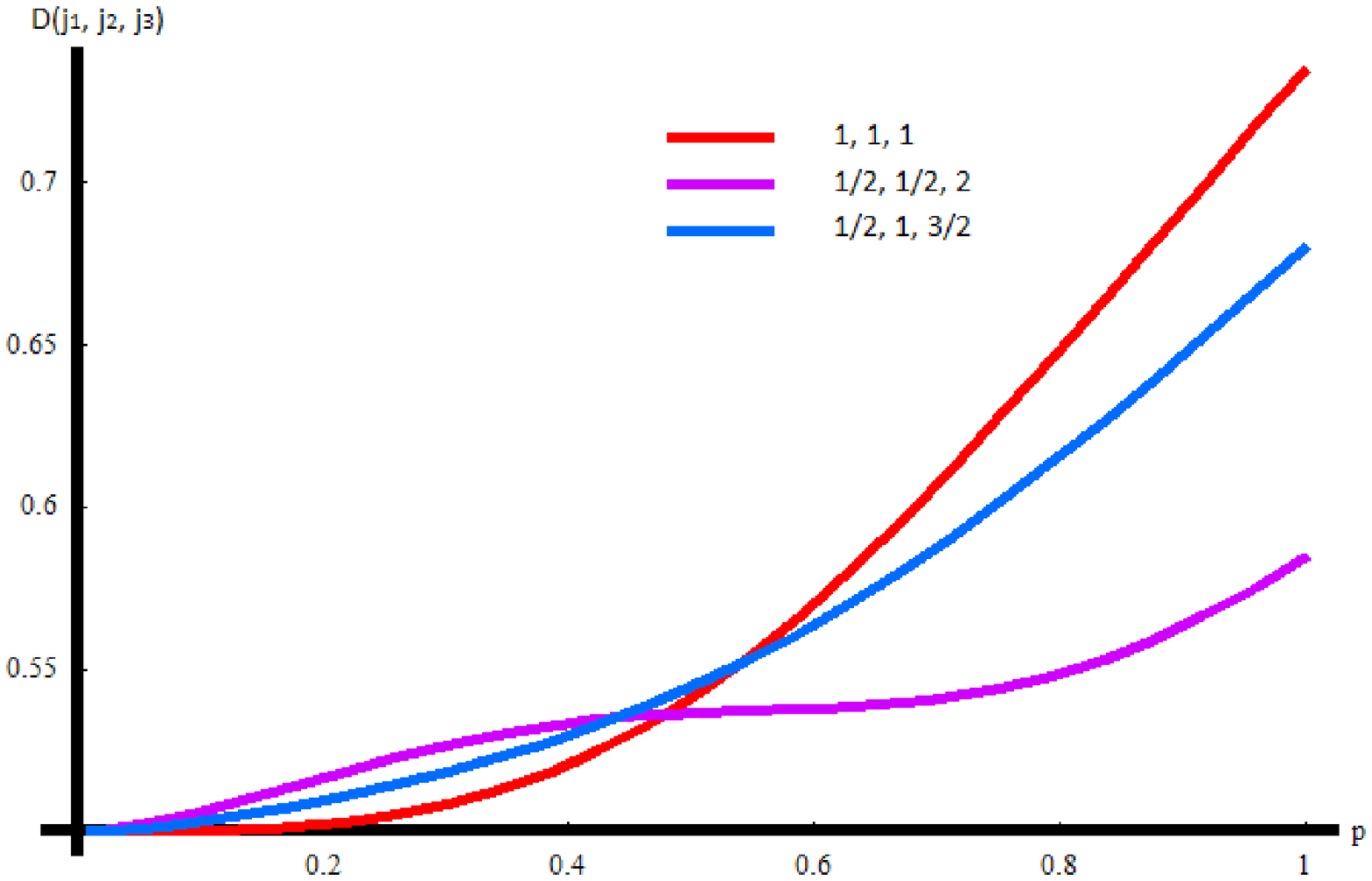}\\
FIG. 7:  {\sf The multipartite quantum correlations for  $j=3$
versus the overlap $p$ for $m=1$.}
\end{center}

From figures 6 and 7, one can see that the tripartite quantum
discord $D(j_1 = 1, j_2 = 1, j_3 = 1)$, $D(j_1 = \frac{1}{2}, j_2 =
\frac{1}{2} , j_3 = 2)$ and $D(j_1 = \frac{1}{2}, j_2 = 1  , j_3 =
\frac{3}{2})$ are all equals for $p \simeq 0.5$. Note also that  for
$p \leq 0.5$,  the sum of all pairwise quantum discord obtained in
the spitting scheme $ (j = 3) \longrightarrow (j_1 = 1, j_2 = 1, j_3
= 1)$ is minimal in comparison with the two others. This behavior
changes  when $p \geq 0.5$ and the  quantity $D(j_1 = 1, j_2 = 1,
j_3 = 1)$ becomes maximal. For even spin coherent states ($m = 0$),
the  measure of tripartite quantum correlations vanishes when $ p
\longrightarrow 1$ as expected (see equations (\ref{Etotal}) and
(\ref{Dtotal})).

\subsection{Monogamy of quantum discord}
In the pure tripartite state (\ref{partition3}), the quantum discord
satisfy the monogamy relation when the following condition
$$D^{\rightarrow}(\rho_{j_1j_2}) + D^{\rightarrow}(\rho_{j_1j_3}) \leq D^{\rightarrow}(\rho_{j_1\vert j_2j_3})$$
is satisfied. As for entanglement of formation, we shall focus on
some special cases to determine the positivity  of the function
$$\Delta D = D^{\rightarrow}(\rho_{j_1\vert j_2j_3})- D^{\rightarrow}(\rho_{j_1j_2}) - D^{\rightarrow}(\rho_{j_1j_3})$$
when the overlap vary from 0 to 1.  We first consider the situation
where $(j_1 = \frac{1}{2},j_2= \frac{1}{2},j_3= \frac{1}{2})$. The
function $\Delta D $ is plotted in figure 8. In this case the
quantum discord is monogamous for even spin coherent state. However,
for odd spin coherent state, the monogamy relation is satisfied only
when $ p \leq 0.8$. We also consider the situations where $(j_1= 1 ,
j_2= \frac{1}{2}, j_3 = \frac{1}{2})$, $(j_1= \frac{1}{2} , j_2= 1,
j_3 = \frac{1}{2})$  and $(j_1= \frac{1}{2}, j_2= \frac{1}{2}, j_3 =
1)$ associated to the spin $j=2$. The behavior of the function
$\Delta D $  for even coherent states $(m = 0)$ is reported in the
figure 9. Clearly, the monogamy relation is satisfied.  The figure
10, representing the function $\Delta D $ for odd case $(m = 1)$,
reveals that the quantum discord ceases to be monogamous for $p$
approaching the unity. Remark that in the figures 9 and 10, we have
$\Delta D ( \frac{1}{2}, 1, \frac{1}{2}) = \Delta D ( \frac{1}{2},
\frac{1}{2}, 1 )$ as expected. It is interesting to note that the
behavior of $\Delta D$ versus $p$ is identical to the ones obtained
for $\Delta E$ in the previous section (figures 3, 4 and 5). This is
essentially due to the conservation relations between quantum
discord and entanglement of formation (\ref{sum-discord}) \cite{Z-H Ma}. Finally,
it is interesting to note that the odd tripartite coherent states
$(m = 1)$ interpolate continuously between the three-qubit
Greenberger-Horne-Zeilinger (${\rm GHZ}_3$) states when $p
\rightarrow 0$ and ${\rm W}_3$ states for $p \rightarrow
1$. It follows from figure 10 that the ${\rm GHZ}_3$ states follow
monogamy and ${\rm W}_3$ states do not.

\begin{center}
\includegraphics[width=4in]{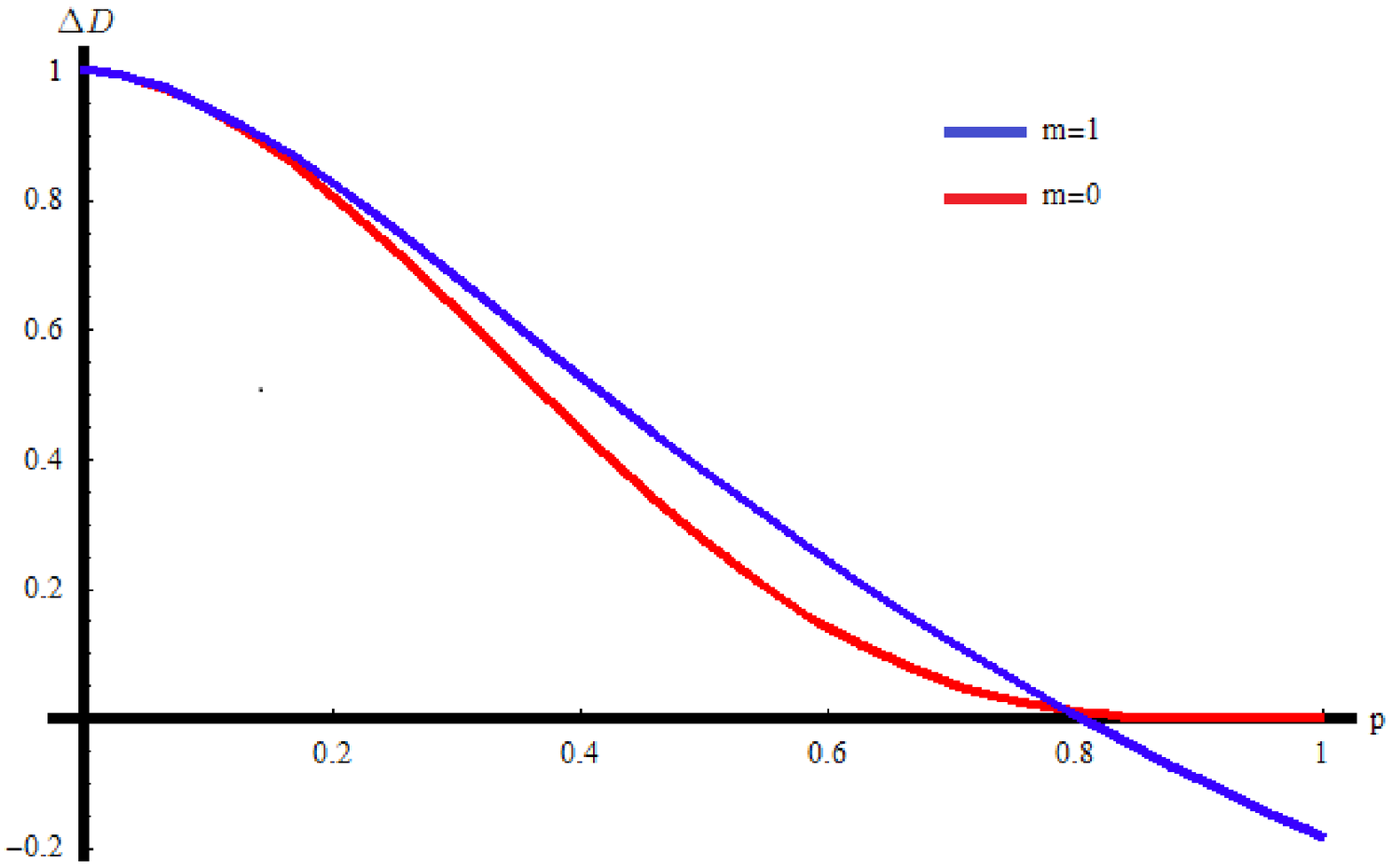}\\
FIG. 8:   {\sf The function $\Delta D$ versus the overlap $p$ when
$j_1=j_2=j_3=\frac{1}{2}$ for $m=0$ and $m=1$.}
\end{center}
\begin{center}
\includegraphics[width=4in]{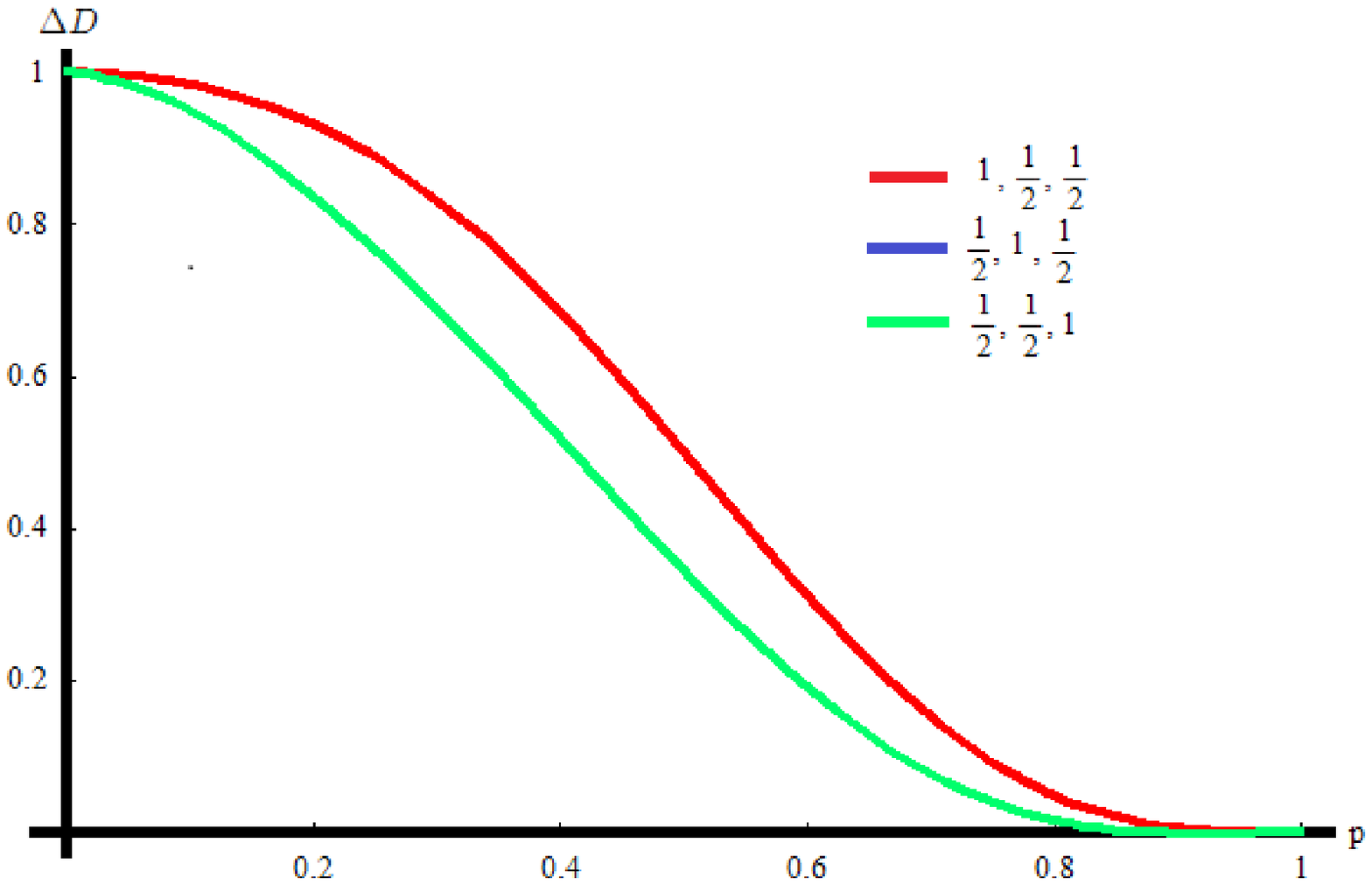}\\
FIG. 9:  {\sf The function $\Delta D$ versus the overlap $p$ when
$(j_1= 1 , j_2= \frac{1}{2}, j_3 = \frac{1}{2})$ and $(j_1=
\frac{1}{2}, j_2= \frac{1}{2}, j_3 = 1)$   for $m=0$.}
\end{center}
\begin{center}
\includegraphics[width=4in]{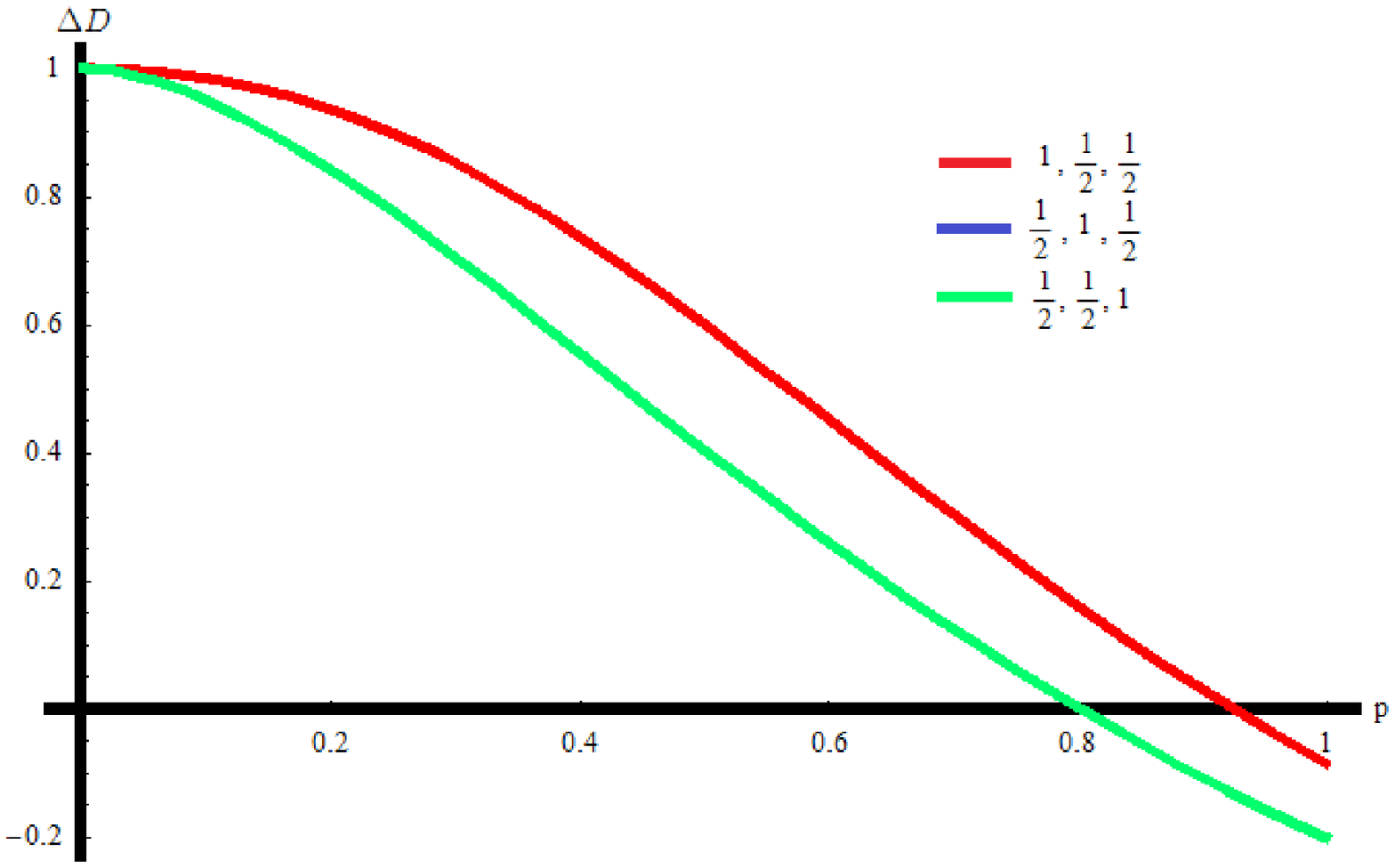}\\
FIG. 10:    {\sf The function $\Delta D$ versus the overlap $p$ when
$(j_1= 1 , j_2= \frac{1}{2}, j_3 = \frac{1}{2})$  and $(j_1=
\frac{1}{2}, j_2= \frac{1}{2}, j_3 = 1)$   for $m=1$.}
\end{center}

\section{ Concluding remarks}

The main motivation in investigating the multipartite quantum
correlations in even and odd coherent states is the decomposition
(or factorization) property given by (\ref{split}). In this way, a
single $j$-spin coherent state is  viewed as comprising two, three
or in general $2j$ qubits. Moreover, this decomposition property allows us
to investigate the  pairwise quantum correlations in a in a single spin
coherent state.  In this paper, we mainly focused on bipartite and
tripartite decomposition. For each case, the spin coherent states
were mapped to two or three qubits system. We have considered the
multipartite quantum correlation in even and odd spin coherent
states measured by entanglement of formation and quantum discord. We
defined the total amount of quantum correlation in spin coherent
states, viewed as multi-components system,  as the sum of all
pairwise quantum correlations. We explicitly derived the expressions
of multipartite entanglement of formation and quantum discord for
even and odd spin coherent states. The sum of all possible pairwise entanglement of formation in an even or odd spin coherent,
viewed as a pure tripartite state, is explicitly derived
and it coincides with sum of pairwise quantum discord of all possible bi-partitions
as it has been shown in \cite{Z-H Ma}. This
peculiar result originates from the conservation relation between
the entanglement of formation and quantum discord given by
(\ref{sum-discord}). We also examined the monogamy relation of
entanglement of formation and quantum discord. Remarkably, in the
simplest cases that we considered, these two measures are monogamous
for even spin coherent contrarily to odd case where the monogamy
relation is violated for states involving an overlap $p$ approaching
the unity. In particular, we have shown that the entanglement of
formation and quantum discord follow the monogamy relation in the
three qubit Greenberger-Horne-Zeilinger states  contrarily  to the
three qubit  states of $W$ type. As prolongation of the present
work, it will be  an important issue to extend the present approach
to others coherent and squeezed states. Further thought in this
direction might be worthwhile in investigating genuine multipartite
quantum correlations. Finally, it is interesting  to examine
the relation between the spin coherent states factorization (\ref{split}) and the
tensor product decomposition of two fermions developed in \cite{Caban}.\\


\begin{thebibliography}{99}

\bibitem{Horodecki}  R. Horodecki, P. Horodecki, M. Horodecki and K. Horodecki, Rev. Mod. Phys. {\bf 81} (2009) 865.

\bibitem{Guhne}  O. G\"uhne and G. T\'oth, Phys. Rep. {\bf 474} (2009) 1.

\bibitem{Modi}  K. Modi, A. Brodutch, H. Cable, T. Paterek and V. Vedral, Rev. Mod. Phys. {\bf 84} (2012) 1655.



\bibitem{Ollivier-PRL88-2001} H. Ollivier and W.H. Zurek, Phys. Rev. Lett. {\bf 88} (2001) 017901.

\bibitem{Vedral-et-al} L. Henderson and V. Vedral, J. Phys. A {\bf 34} (2001)  6899.

\bibitem{Luo} S. Luo, Phys. Rev. A \textbf{77} (2008) 042303;  Phys.
Rev. A \textbf{77} (2008) 022301.

\bibitem{Ali} M. Ali, A.R.P. Rau and G. Alber,  Phys. Rev. A {\bf 81} (2010) 042105.

\bibitem{Shi1} M. Shi, W. Yang, F. Jiang and J. Du, J. Phys. A: Math. Theor.
{\bf 44} (2011) 415304.

\bibitem{Girolami} D. Girolami and G. Adesso, Phys. Rev. A {\bf 83} (2011)
052108.

\bibitem{Shi2} M. Shi, F. Jiang, C. Sun and J. Du, New J. Phys.  {\bf 13} (2011)
073016.

\bibitem{Rachid1}  M. Daoud and R. Ahl Laamara, J. Phys. A: Math. Theor. {\bf 45} (2012) 325302.

\bibitem{Rachid2}  M. Daoud and R. Ahl Laamara,
Int. J. Quantum Inform. {\bf 10} (2012)
1250060.

\bibitem{Dakic2010} B. Dakic, V. Vedral and C. Brukner, Phys. Rev.
Lett. {\bf 105} (2010) 190502.


\bibitem{Wootters98} C.H. Bennett, D.P. DiVincenzo, J. Smolin and W.K. Wootters, Phys. Rev. A {\bf 54}  (1997) 3814.


\bibitem{Sanders} B.C. Sanders, Phys. Rev. A {\bf 45} (1992) 6811.

\bibitem{Sanders2}  B.C. Sanders, Phys. Rev. A {\bf 46} (1992) 2966.

\bibitem{Sanders3} B.C Sanders, J. Phys. A: Math. Theor. {\bf 45} (2012) 244002.

\bibitem{Crypto2}  C.A. Fuchs, Phys. Rev. Lett. {\bf 79} (1997) 1162.

\bibitem{Qip}  H. Jeong, M.S. Kim and J. Lee, Phys. Rev. A {\bf 64},
 (2001) 052308.

\bibitem{Bartlett}  S.D. Bartlett, H. de Guise and B.C. Sanders, Phys. Rev. A {\bf 65} (2002) 052316.

\bibitem{Jeong}  H. Jeong and M.S. Kim, Phys. Rev. A {\bf 65} (2002) 042305.

\bibitem{Ralph}  T.C. Ralph, W.J. Munro and G.J. Milburn,
Phys. Rev. A {\bf 68}(2003) 042319 .

\bibitem{Lloyd99}  S. Lloyd and S.L. Braunstein, Phys. Rev. Lett. {\bf 82} (1999) 1784.

\bibitem{Cochrane}  P.T. Cochrane, G.J. Milburn and W.J. Munro, Phys.
Rev. A {\bf 59} (1999) 2631.

\bibitem{Oliveira}  M.C. de Oliveira and W.J. Munro, Phys. Rev. A
{\bf 61} (2000) 042309.




\bibitem{Klyachko1} M.A. Can  , A. Klyachko and A. Shumovsky,
J. Opt. B: Quantum Semiclass. Opt. {\bf 7}
 (2005) L1.



\bibitem{Klyachko2} S. Binicioglu, M.A. Can, A.A. Klyachko and  A.S. Shumovsky,
Found. Phys. {\bf 37} (2007)  1253.


\bibitem{Terra} M.O. Terra Cunha, J.A. Dunningham and V. Vedral, Proc. R. Soc. A
{\bf 463} (2007) 2277.

\bibitem{Coffman}  V. Coffman, J. Kundu and W.K. Wootters, Phys. Rev.
A {\bf 61} (2000) 052306.

\bibitem{Adesso2} G. Adesso and F. Illuminati, New J. Phys. {\bf 8} (2006)
15.

\bibitem{Adesso3} T. Hiroshima, G. Adesso and F. Illuminati, Phys. Rev.
Lett. {\bf 98} (2007) 050503.


\bibitem{Giorgi} G.L. Giorgi, Phys. Rev. A {\bf 84} (2011) 054301.

\bibitem{Prabhu}  R. Prabhu, A.K. Pati, A.S. De and U. Sen, Phys. Rev. A
{\bf 86} (2012) 052337.

\bibitem{Sudha} Sudha, A.R. Usha Devi and A.K. Rajagopal, Phys. Rev. A {\bf 85}
(2012) 012103.

\bibitem{Allegra}  M. Allegra, P. Giorda   and A. Montorsi, Phys. Rev. B {\bf 84}
(2011) 245133.

\bibitem{Ren} X.-J. Ren and H. Fan, Quant. Inf. Comp.  {\bf 13} (2013) 0469.

\bibitem{Bruss} A. Streltsov, G. Adesso, M. Piani and D. Bruss, Phys. Rev.
Lett. {\bf 109} (2012) 050503.

\bibitem{Z-H Ma} Z-H Ma, Z-H Chen and F.F. Fanchini, New J. Phys.  {\bf 15}
(2013) 043023.

\bibitem{Fanchini1}  F.F. Fanchini, M.F. Cornelio, M.C. de Oliveira and A.O. Caldeira, Phys. Rev. A {\bf 84} (2011) 012313.

\bibitem {Luo08} S. Luo, Phys. Rev. A \textbf{77} (2008) 042303; Phys.
Rev. A \textbf{77} (2008) 022301.


\bibitem{Adesso1} G. Adesso and A. Datta, Phys. Rev. Lett. {\bf 105}  (2010)
030501;  G. Adesso and D. Girolami, Int. J. Quantum Inform. {\bf 9}
(2011) 1773.

\bibitem{Rachid3} M. Daoud and R. Ahl Laamara,
 Phys. Lett. A {\bf 376} (2012) 2361.

\bibitem{Koachi-Winter} M. Koachi and A. Winter, Phys. Rev. A {\bf
69} (2004) 022309.

\bibitem{Shi} M. Shi, W. Yang, F. Jiang and J. Du,  J. Phys. A: Math. Theor. {\bf 44} (2011)
415304.

\bibitem{Hil97} S. Hill and W.K. Wootters, Phys. Rev. Lett. {\bf 78} (1997)
5022.

\bibitem{Okrasa} M. Okrasa and Z. Walczak, Eur. Phys. Lett. {\bf 96} (2011) 60003.

\bibitem{Chakrabarty} I. Chakrabarty, P. Agrawal and A.K. Pati, Eur. Phys. J. D  {\bf 65} (2011) 605.


\bibitem{Rulli} C.C. Rulli and M.S. Sarandy, Phys. Rev. A {\bf 84} (2011) 042109.

\bibitem{Fanchini2}  F.F. Fanchini, M.C. de Oliveira, L.K. Castelano  and M.F. Cornelio, Phys. Rev. A {\bf 87} (2013) 032317.

\bibitem{GHZ} D.M. Greenberger, M.A. Horne and A. Zeilinger, Physics Today
{\bf 46} (1993) 22.

\bibitem{Dur00} W. D\"ur, G. Vidal and J.I. Cirac, Phys. Rev. A
{\bf 62} (2000) 062314.

\bibitem{Fanchini} F.F. Fanchini, L.K. Castelano, M.F. Cornelio and M.C. de Oliveira,  New J. Phys.  {\bf 14} (2012)
013027.

\bibitem{Caban} P. Caban, K. Podlaski, J. Rembieli\'nski, K. A. Smoli\'nski
and Z. Walczak, J. Phys. A: Math. Gen. {\bf 38} (2005) L79.

















\end{thebibliography}
\end{document}